\documentstyle[11pt,amssym,emulateapj_rj,epsfig]{article}

\topline{The Astrophysical Journal Supplement Series, 127:000--000, 2000
         {\rm March}\hspace*{6cm}(MS-40536-ApJS)}
\submitted{Received 1999 May 5; accepted 1999 August 19}
\received{May 5, 1999}
\accepted{August 19, 1999}
\journalid{127}{2000 March \#1 issue}
\articleid{000}{000}
\paperid{MS-40536-ApJS}

\lefthead{JANSEN $ET\, AL$.}
\righthead{SURFACE PHOTOMETRY OF NEARBY FIELD GALAXIES}

\newcommand{\ie}{\mbox{i.e.\ }}%
\newcommand{\eg}{\mbox{e.g.\ }}%
\newcommand{\cf}{\mbox{cf.\ }}%
\newcommand{\etal}{\mbox{et al. }}%
\newcommand{\tsim}{\mbox{$\sim$}}%
\newcommand{\tpm}{\mbox{$\pm$}}%
\newcommand{\muR}{\mbox{$\mu_{{}_R}$}}%
\newcommand{\muB}{\mbox{$\mu_{{}_B}$}}%
\newcommand{\muU}{\mbox{$\mu_{{}_U}$}}%
\newcommand{\muBe}{\mbox{$\mu^B_e$}}%
\newcommand{\rBe}{\mbox{$r^B_e$}}%
\newcommand{\MB}{\mbox{M$_B$}}%
\newcommand{\MZ}{\mbox{M$_Z$}}%
\newcommand{\BR}{\mbox{$(B\!-\!R)$}}%
\newcommand{\br}{\mbox{$B\!-\!R$}}%
\newcommand{\BRe}{\mbox{\BR$_e$}}%
\newcommand{\UB}{\mbox{$(U\!-\!B)$}}%
\newcommand{\UBe}{\mbox{\UB$_e$}}%
\newcommand{\UR}{\mbox{$(U\!-\!R)$}}%
\newcommand{\persecsq}{\mbox{$\;$arcsec$^{-2}$}}%
\newcommand{\magpssq}{\mbox{mag/$\Box ^2$}}%
\newcommand{\kms}{\mbox{km s$^{-1}$}}%
\newcommand{\Ho}{\mbox{H$_0$}}%
\newcommand{\kmsMpc}{\mbox{km s$^{-1}$ Mpc$^{-1}$}}%
\newcommand{\HI}{\mbox{H{\sc i}$\;$}}%
\newcommand{\HII}{\mbox{H{\sc ii}$\;$}}%
\newcommand{\lii}{\mbox{$l_{\hbox{\sc ii}}$}}%
\newcommand{\bii}{\mbox{$b_{\hbox{\sc ii}}$}}%
\newcommand{\vlg}{\mbox{$V_{LG}$}}%

\begin{document}

\title {\large\bf Surface photometry of nearby field galaxies: the data}

\author{Rolf A. Jansen$^{1,2}$, Marijn Franx$^{1,3}$, 
        Daniel Fabricant$^{2}$, \& Nelson Caldwell$^{4,2}$}
\affil {$^1$Kapteyn Astronomical Institute, Postbus~800, NL-9700~AV
	Groningen, The Netherlands}
\affil {$^2$Harvard-Smithsonian Center for Astrophysics, 60 Garden St.,
	Cambridge, MA~02138}
\affil {$^3$Leiden Observatory, Postbus 9513, NL-2300 RA Leiden,
        The Netherlands}
\affil {$^4$F.L.~Whipple Observatory, Amado, AZ~85645}
\affil {jansen@astro.rug.nl, franx@strw.leidenuniv.nl,
	dfabricant or ncaldwell@cfa.harvard.edu}

%%%%%%%%%%%%%%%%%%%%%%%%%%%%%%%%%%%%%%%%%%%%%%%%%%%%%%%%%%%%%%%%%%%%%%%%
% ABSTRACT  ABSTRACT  ABSTRACT  ABSTRACT  ABSTRACT  ABSTRACT  ABSTRACT %
%%%%%%%%%%%%%%%%%%%%%%%%%%%%%%%%%%%%%%%%%%%%%%%%%%%%%%%%%%%%%%%%%%%%%%%%

\begin{abstract}

We have obtained integrated spectra and multi-filter photometry for a
representative sample of \tsim 200 nearby galaxies.  These galaxies span
the entire Hubble sequence in morphological type, as well as a wide
range of luminosities ($\MB=-14$ to $-22$) and colors ($\br=0.4$ to
1.8).  Here we describe the sample selection criteria and the $U, B, R$
surface photometry for these galaxies.  The spectrophotometric results
will be presented in a companion paper.  Our goals for the project
include measuring the current star formation rates and metallicity of
these galaxies, and elucidating their star formation histories, as a
function of luminosity and morphology.  We thereby extend the work of
Kennicutt (1992) to lower luminosity systems.  We anticipate that our
study will be useful as a benchmark for studies of galaxies at high
redshift. 

We discuss the observing, data reduction and calibration techniques, and
show that our photometry agrees well with previous work in those cases
where earlier data are available.  We present an atlas of images, radial
surface brightness profiles and color profiles, as well as tables of
derived parameters.  The atlas and tables of measurements will be made
available electronically. 

We study the correlations of galaxy properties determined from the
galaxy images.  Our findings include: (1) colors determined within the
effective radius correlate better with morphological type than with \MB\
and (2) 50\% of the low luminosity galaxies are bluest in their centers. 

\end{abstract}


%%%%%%%%%%%%%%%%%%%%%%%%%%%%%%%%%%%%%%%%%%%%%%%%%%%%%%%%%%%%%%%%%%%%%%%%
% INTRODUCTION  INTRODUCTION  INTRODUCTION  INTRODUCTION  INTRODUCTION %
%%%%%%%%%%%%%%%%%%%%%%%%%%%%%%%%%%%%%%%%%%%%%%%%%%%%%%%%%%%%%%%%%%%%%%%%

\section{Introduction}

With the advent of the Hubble Space Telescope, 8-m class ground-based
telescopes, and effective new instruments, it is now possible to study
the colors, morphologies and spectra of large samples of distant, faint
galaxies.  Frequently, our interpretation of such data is limited by the
scarcity of good comparison samples in the local universe.  Because
distant galaxies subtend angles comparable to spectrograph slit widths,
we usually obtain integrated spectra, while the same slits sample only
the nuclear regions of nearby galaxies.  This observational issue
complicates the comparison of distant and nearby galaxy spectra. 

In a pioneering effort, Kennicutt (1992ab) obtained integrated
spectrophotometry for 90 galaxies spanning the entire Hubble sequence. 
This study has had broad application for the study of galaxy spectral
properties at both high and low redshift.  Kennicutt's study, however,
is limited to the brightest galaxies of each morphological type, and no
uniform, multiple filter, surface photometry is available for these
galaxies.  Also, only half of Kennicutt's sample of 90 galaxies was
observed at 5--7\AA\ spectral resolution by trailing the image of a
galaxy across a long slit.  The remaining half was observed at
15--20\AA\ resolution through a 45.5\arcsec\ circular aperture. 

With our Nearby Field Galaxy Survey (NFGS), our goal is to significantly
extend Kennicutt's work.  We have obtained integrated and nuclear
spectroscopy, and $U,B,R$ surface photometry, for a sample of 196
galaxies in the nearby field.  This sample includes galaxies of all
morphological types and spans 8 magnitudes in luminosity.  We include
galaxies from a broad range of local galaxy densities, attempting to
avoid a bias towards a particular environment.  Our use of the term
``field'' corresponds to that of Koo \& Kron (1992) and Ellis (1997). 
We use these observations to study the emission and absorption line
strengths, metallicities, star formation rates and star formation
histories, morphologies, structural parameters, and colors of the sample
galaxies.  These data can be used as an aid in understanding the spectra
and imagery of galaxies at larger distances, and in measuring the
changes in their properties over time. 

In this first paper we describe the sample selection and present the
surface photometry data.  In subsequent papers we will present the
spectrophotometric data, and present further analysis of the complete
data set.  The structure of this paper is as follows. 
Section~\ref{P-samplesel} discusses the sample selection.  In
Section~\ref{P-obsredcal} we discuss the observing strategy, data
reduction, calibration issues and an assessment of the data quality and
errors.  We present the primary data products in
section~\ref{P-results}.  We conclude with a short discussion
(section~\ref{P-discussion}).  Notes on individual objects in the sample
are collected in appendix~\ref{P-App.objects}, and the details of the
adopted photometric calibration procedure are given in
appendix~\ref{P-App.calib}.

%%%%%%%%%%%%%%%%%%%%%%%%%%%%%%%%%%%%%%%%%%%%%%%%%%%%%%%%%%%%%%%%%%%%%%%%
% SECTION  SECTION  SECTION  SECTION  SECTION  SECTION  SECTION  SECTION
%%%%%%%%%%%%%%%%%%%%%%%%%%%%%%%%%%%%%%%%%%%%%%%%%%%%%%%%%%%%%%%%%%%%%%%%

\section{Sample selection}
\label{P-samplesel}

We selected galaxies from the first CfA redshift catalog (CfA~I), which
contains galaxies to a limiting blue photographic magnitude of
$m_Z=14.5$ (Huchra \etal 1983).  This catalog has several virtues: (1)
it is nearly complete within its selection limits, (2) it contains
galaxies with a large range in absolute magnitude ($-22\lesssim
\MZ\lesssim -13$ for \Ho=100 \kmsMpc), and (3) all galaxies in the
catalog have been morphologically classified.  As the CfA~I catalog is
too large (\tsim 2400 galaxies) for a full spectroscopic and photometric
survey, we have selected a subsample of \tsim 200 galaxies as described
below.  Our goal was to select a sample of galaxies which spans the full
range of galaxy properties in the CfA sample. 

Because the FAST spectrograph (Fabricant \etal 1998) used to obtain the
galaxy spectra for this survey has a maximum slit length of \tsim
3\arcmin, we wished to minimize the number of galaxies larger than the
slit length.  We avoided a strict diameter limit that might introduce a
bias against low surface brightness galaxies.  Instead, we selected
higher luminosity galaxies (generally the largest) at greater distances
by imposing a lower limit on radial velocity that increases with
luminosity.  After applying a correction for the motion of the Milky Way
with respect to the Local Group standard of rest, this cutoff is \vlg
(\kms) $> 10^{-0.19 - 0.2 \hbox{\scriptsize M}_Z}$.  We take for the
velocity with respect to the Local Group standard of rest: $\vlg =
V_{helio} + 300\,cos(b)sin(l)$.  This approximates the result of Yahil
\etal (1977).  Here, we calculate galaxy distances directly from \vlg,
assuming \Ho=100 \kmsMpc.  No correction for infall towards the Virgo
Cluster was applied at this stage. 

In order to avoid a sampling bias favoring a cluster population we also
excluded galaxies in the direction of the Virgo Cluster with $\vlg <
2000$ \kms\ in the RA and DEC ranges given by Binggeli \etal (1985).  A
total of 1006 galaxies are left after our radial velocity and Virgo
Cluster cuts, only a few of which are larger than 3\arcmin. 

Our goal was to reduce this number to \tsim 200 galaxies with a
distribution in \MZ\ that spans the full range in galaxy luminosities,
and that fairly samples the changing mix of morphological types as a
function of \MZ.  To this end, the sample was sorted into 1 magnitude
wide bins of \MZ\ and within each bin by morphological type.  We then
selected every $N$th galaxy in each bin, where $N$ is the ratio between
the total and desired number of galaxies in the bin (chosen to
approximate the local galaxy luminosity function, \eg~Marzke \etal
1994).  The total number of galaxies entering our statistical sample is
196.  The median redshift in this sample is 0.01, and the maximum
redshift is \tsim 0.07 (A12195\-$+$7535). 

The morphological types used in the selection were taken from Huchra
\etal (1983).  The types as used elsewhere in this paper are mainly from
a later release of the CfA redshift survey and its extensions (Huchra
\etal 1997) and from independent reclassifications by the authors,
S.~Kannappan and J.P.~Huchra, using our CCD imagery.  The main
difference between the original and updated types is that the fraction
of unclassified spirals and ``peculiars'' has been reduced, and that a
few galaxies that were clearly mistyped have been reclassified.  The
reclassifications have not biased the sample, as they are redistributed
fairly evenly over morphological type. 

In table~1 we collect the global parameters of our sample as used in the
sample selection.  Column (1) lists the galaxy identifications as used
throughout this paper, while columns (2) and (3) give the common names
(NGC, IC, or IAU anonymous notation) and UGC catalog number (Nilson
1973), respectively.  Columns (4) through (7) contain the equatorial and
Galactic coordinates.  The equatorial coordinates were remeasured from
the Digital Sky Survey (DSS), reduced to B1950.0, and are accurate to
\tsim 1\arcsec.  Columns (8) through (10) list the original numeric
morphological types (Huchra \etal 1983) as used in the sample selection,
and the numeric types and their translation on the Hubble sequence as
adopted throughout the further analysis of the data.  Columns (11) and
(12) give the systemic velocities, both heliocentric and reduced to the
Local Group standard of rest.  Column (13) lists the absolute magnitude
calculated from the blue photographic magnitude (CGCG, Zwicky \etal
1961--8) and the systemic velocity, assuming a simple uniform Hubble
flow and \Ho=100 \kmsMpc.  Co\-lumns (14) and (15) list the blue \vfill

%%%%%%%%%%%%%%%%%%%%%%%%%%%%%%%%%%%%%%%%%%%%%%%%%%%%%%%%%%%%%%%%%%%%%%%%%
%                                                                       %
%                        PHOTOMETRY DATA PAPER                          %
%	     TABLE 1: SAMPLE DEFINITION / GLOBAL PARAMETERS		%
%                                                                       %
%%%%%%%%%%%%%%%%%%%%%%%%%%%%%%%%%%%%%%%%%%%%%%%%%%%%%%%%%%%%%%%%%%%%%%%%%
%\landscape
\setlength{\tabcolsep}{1pt}


%\endlandscape
\twocolumn\clearpage
\vfill

%%%%%%%%%%%%%%%%%%%%%%%%%%%%%%%%%%%%%%%%%%%%%%%%%%%%%%%%%%%%%%%%%%%%%%%%%
%         FIGURE 1: OVERVIEW GLOBAL PROPERTIES SELECTED SAMPLE		%
%%%%%%%%%%%%%%%%%%%%%%%%%%%%%%%%%%%%%%%%%%%%%%%%%%%%%%%%%%%%%%%%%%%%%%%%%
\newlength{\txw}
\setlength{\txw}{\textwidth}

\noindent\leavevmode
\makebox[\txw]{
   \centerline{
      \epsfig{file=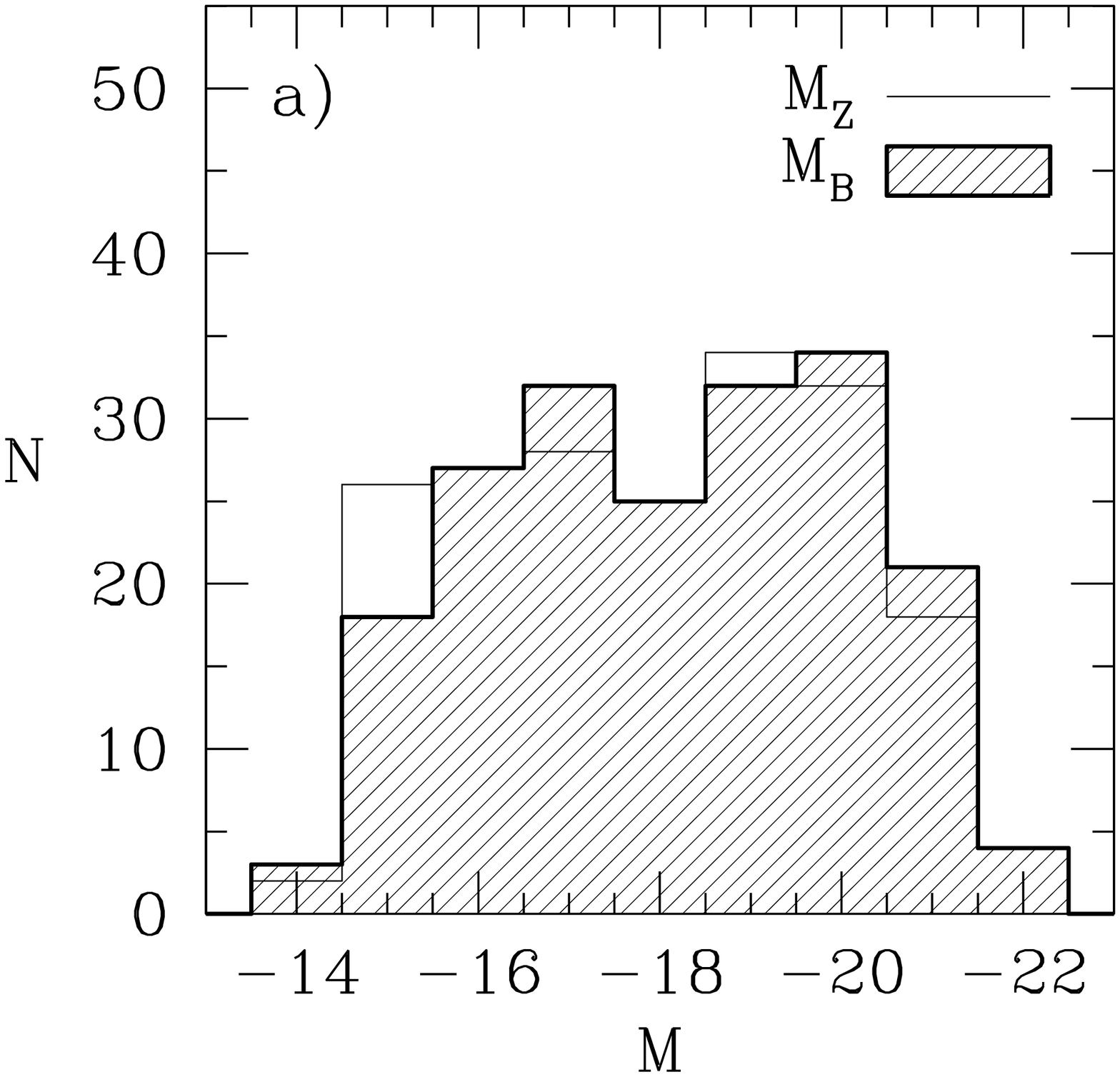,width=0.33\txw,clip=}
      \epsfig{file=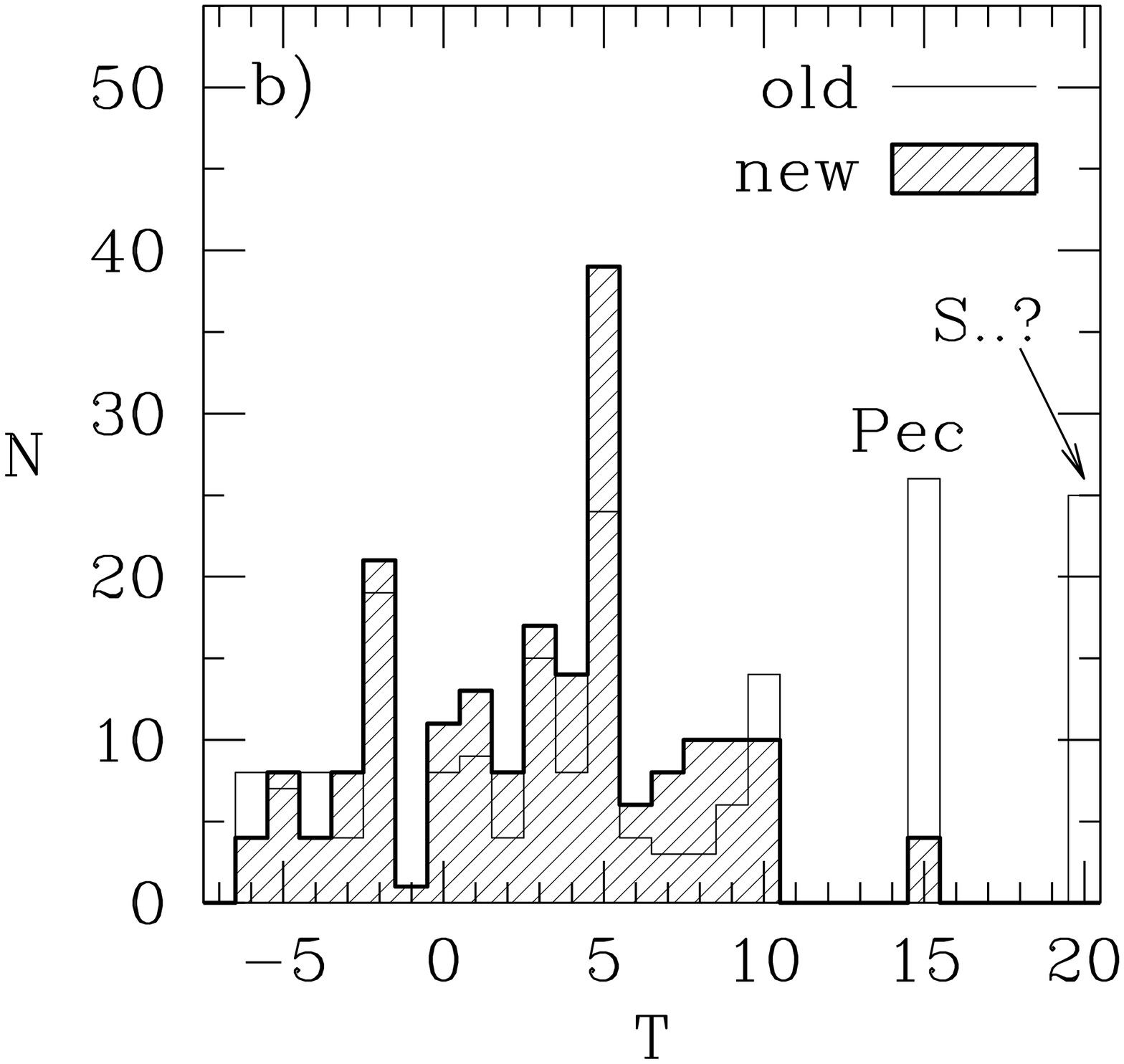,width=0.33\txw,clip=}
      \epsfig{file=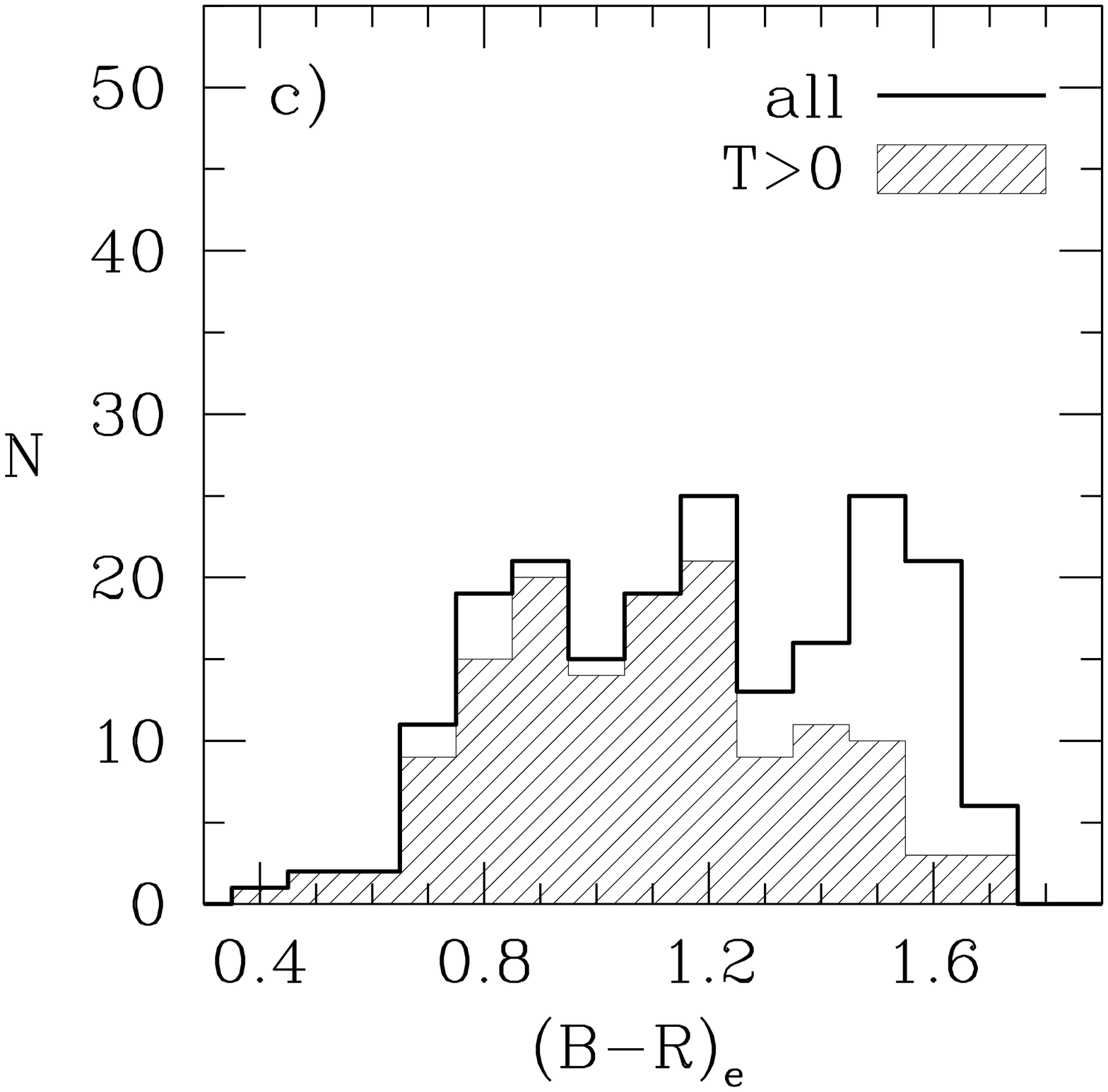,width=0.33\txw,clip=}
   }
}\par\noindent\leavevmode
\makebox[\txw]{
   \centerline{
      \epsfig{file=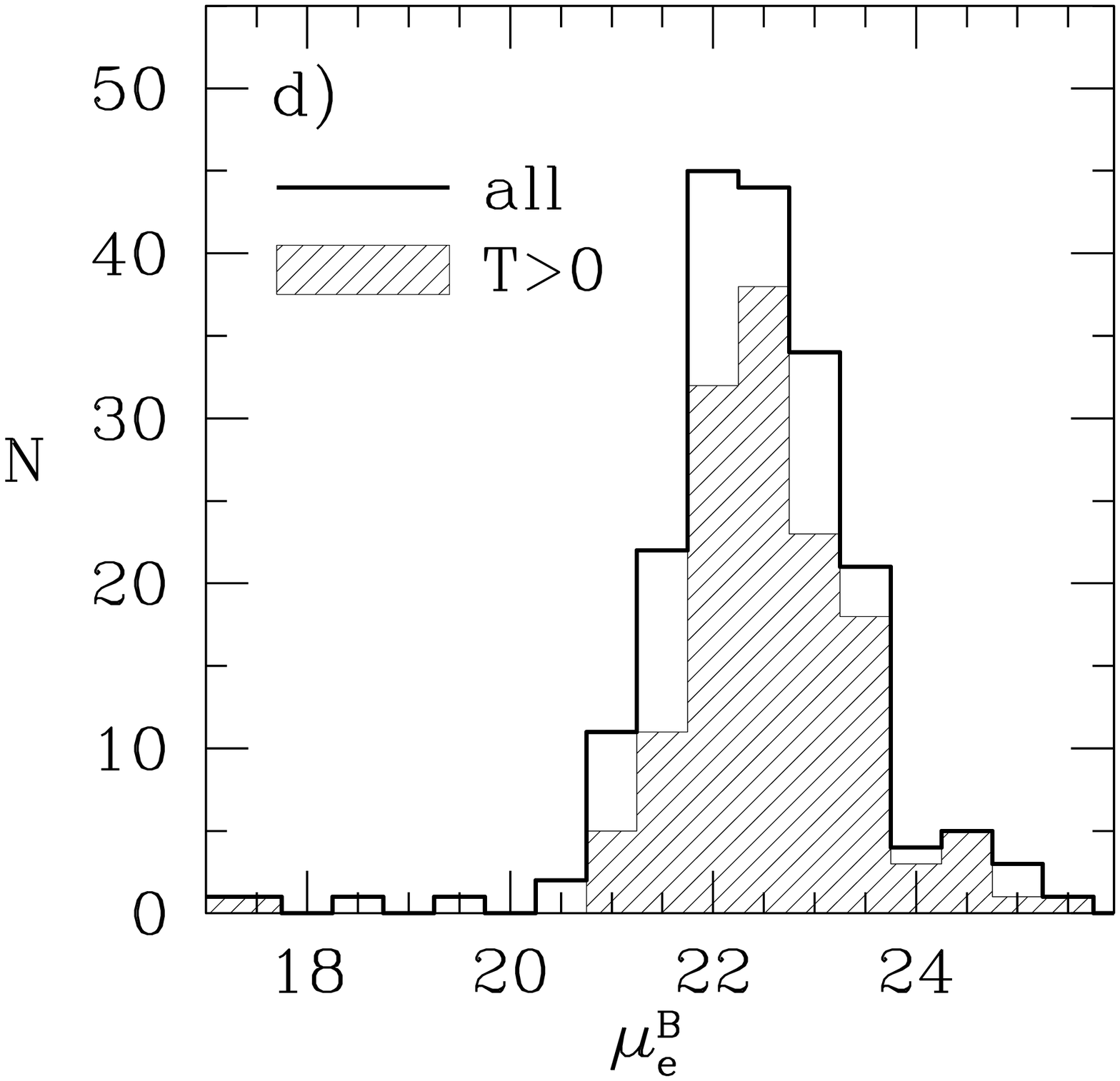,width=0.33\txw,clip=}
      \epsfig{file=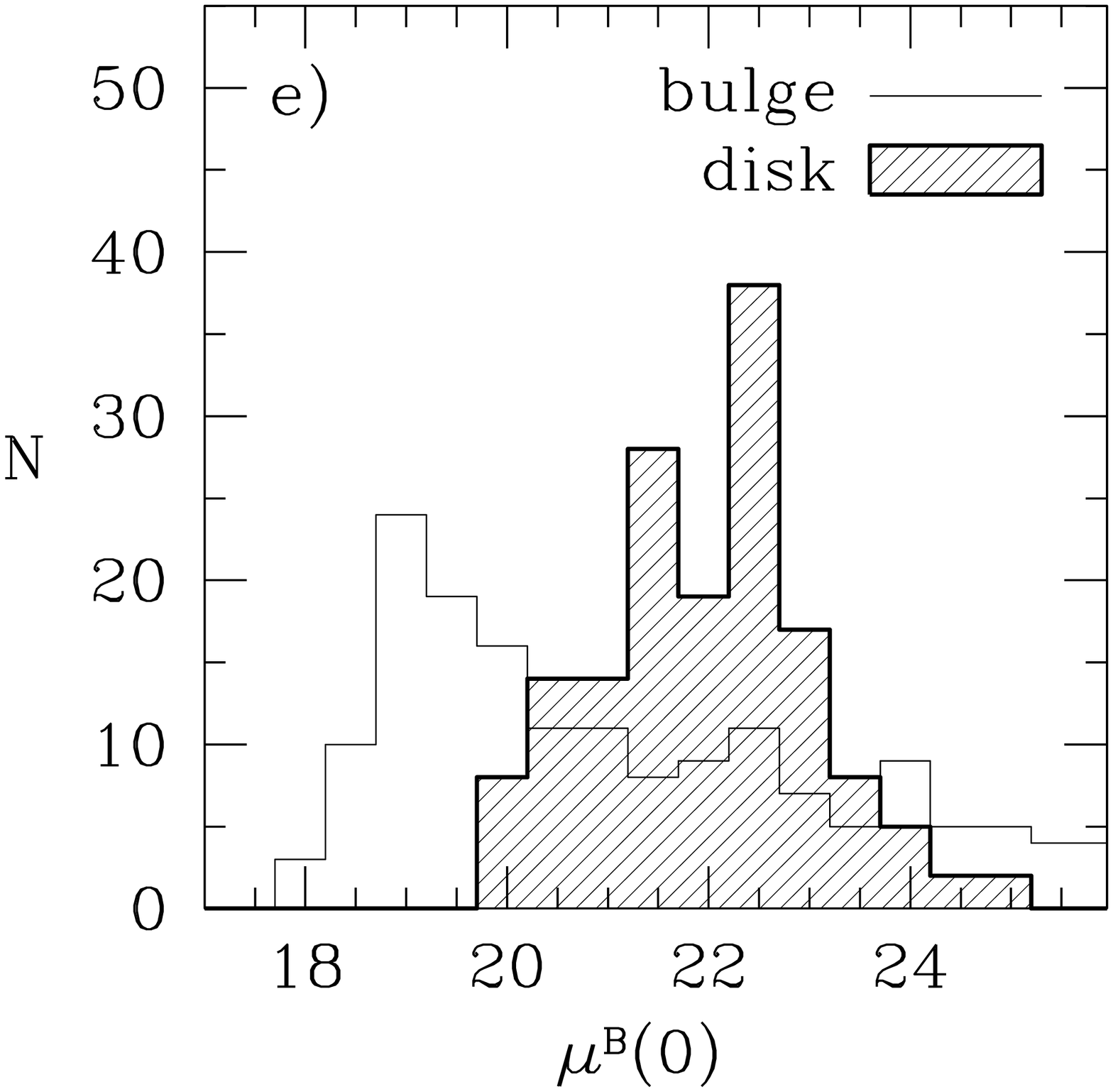,width=0.33\txw,clip=}     
      \makebox[0.33\txw]{\rule{0pt}{0.325\txw}}
   }
}\par\noindent\makebox[\txw]{
\centerline{
\parbox[t]{\txw}{\footnotesize {\sc Fig.~1 ---} Overview of the global
properties of the selected galaxy sample.  We present the number
distributions as a function of a) absolute Zwicky (open) and $B$
(shaded) magnitudes; b) galaxy type, for the original type used in the
selection of the sample (open) and for the types adopted in the present
analysis (shaded).  Note that here and in the following figures we place
the unclassified early type galaxies at T=$-4$, as this bin is empty (we
have no cD galaxies in our sample) and the proper placement at T=$-7$
seems awkward with respect to the compact ellipticals at T=$-6$; c)
effective color \BRe, measured within the effective radius in $B$.  d)
the surface brightness at the effective radius in $B$.  In panels c) and
d) the open histograms represent the selected sample, and the shaded
histograms the subsample of galaxies later than S0/a; e) the face-on
(extrapolated) central surface brightness of disks and
spheroidal/elliptical components.  } }
}\vspace*{0.5cm}
%
%%%%%%%%%%%%%%%%%%%%%%%%%%%%%%%%%%%%%%%%%%%%%%%%%%%%%%%%%%%%%%%%%%%%%%%%
\noindent\leavevmode\makebox[\txw]{
   \parbox[b]{0.475\txw}{
\noindent major axis diameters, $a$, and the ellipticities,
$\epsilon=1-b/a$, derived from the minor and major axis ratio as listed
in the UGC catalog.  Column (16) gives the position angles of the major
axis, measured from North through East, as given in the UGC.  These
position angles and the ellipticities in column (15) were used
(section~\ref{P-reduction}) to obtain radial intensity profiles.  Column
(17) gives the $B$-filter Galactic foreground extinction as estimated
from the \HI maps of Burstein \& Heiles (1984) as given in the RC3
(de~Vaucouleurs \etal 1991).  The final column lists the available data,
where abbreviations ``i'', ``n'', ``u'', ``b'', and ``r'' are used for
integrated and nuclear spectra, and $U,B$ and $R$ surface photometry,
respectively.  Italic font is used where no photometric exposures are
available in $U,B$, or $R$; bold-face font where the only observation
was a short photometric ``snapshot'' exposure (see
section~\ref{P-observations}). }
   \parbox[b]{0.040\txw}{\rule{0pt}{0.1\txw}}
   \parbox[b]{0.475\txw}{
\noindent To efficiently use the available telescope time, several more
galaxies were included in our observations. The data for two of these,
A01047\-$+$1625 and NGC~784, both very late type and very nearby, are
presented in this paper as well, but are not part of the statistical
sample.  In table~1 these galaxies are shown with parentheses. 

In figure~1 we give an overview of the main sample properties, comparing
the selection properties with the observed magnitudes, colors and types. 
In figures~1{\em a--d}\ we plot for the 196 sample galaxies the number
distributions of absolute magnitude (\MB\ and \MZ), morphological type
(as selected and as reclassified), effective color \BRe, measured within
the effective radius \rBe\ in $B$, and surface brightness in $B$ at that
effective radius. 

We prefer the surface brightness measured at the effective radius over
the (extrapolated) central surface}

}
\vfill\newpage

\noindent brightness, because (1) it can be measured for any light
distribution (we need not a priori adopt a model for the light
distribution, which might introduce artificial differences between
galaxies of different morphological type), (2) the measurement is
relatively robust, (3) no extrapolation is involved, and (4) it can be
readily compared with results from numerical galaxy simulations. 
However, to facilitate comparison with data in the literature, we also
fit the face-on (extrapolated) central disk surface brightness and
exponential scale length for the galaxies with a disk component and,
after subtraction of the disk, the face-on central surface brightness of
bulge, or central condensation.  The bulge and elliptical central
surface brightnesses are uncertain as they depend critically on the
light distribution in the inner few arcseconds of a galaxy.  The result
is plotted in figure~1{\em e}.  We note that for a pure exponential disk
the central surface brightness and the surface brightness at the
effective radius are related by $\muBe = \mu^B(0) +
1.0857\,(\rBe/r^B_0)+2.5\,\log(1-\epsilon)$, with $r^B_0$ the
exponential scale length.

%%%%%%%%%%%%%%%%%%%%%%%%%% PLACE FIGURE HERE %%%%%%%%%%%%%%%%%%%%%%%%%%%
%
\placefigure{P-Fig.sampledef}
%
%%%%%%%%%%%%%%%%%%%%%%%%%%%%%%%%%%%%%%%%%%%%%%%%%%%%%%%%%%%%%%%%%%%%%%%%

%%%%%%%%%%%%%%%%%%%%% PLACE TABLE IN THIS SECTION %%%%%%%%%%%%%%%%%%%%%%
%
%\placetable{P-Tab.sampledef}
%\begin{table}\dummytable\label{P-Tab.sampledef}\end{table}
%
%%%%%%%%%%%%%%%%%%%%%%%%%%%%%%%%%%%%%%%%%%%%%%%%%%%%%%%%%%%%%%%%%%%%%%%%

\subsection*{Merits and limitations of the sample}
\label{P-merits}

The final sample contains 196 nearby galaxies with a large range of
absolute magnitudes ($-14<\MB<-22$), extending about five magnitudes
fainter than the characteristic local galaxy luminosity in $B$ ($M_*\sim
-19.2$, de Lapparent \etal 1989; $M_*\sim -18.8$, Marzke \etal 1994). 
By the selection process we assure sampling of the entire range of
morphologies.  Galaxies with several types of nuclear activity (Seyfert
{\sc i} and Seyfert {\sc ii}; BL~Lac; nuclear star burst) are present in
our sample.  We also fairly sample a range of \tsim 1.3 magnitude in
effective \BRe\ color and the number distribution as a function of color
is broad.  (Our selection was unbiased towards color.) As expected, the
early type galaxies are predominantly red while the spirals and later
type galaxies cover a large range in color.  The distribution of
observed surface brightness at the effective radius in $B$ has a peak at
22.0 mag\persecsq, a mean of 22.4 mag\persecsq, and a FWHM of 1.7
mag\persecsq.  The tails of the distribution, however, extend over
almost 10 mag\persecsq.  The early type galaxies tend to be brighter at
the effective radius than the late type galaxies.  The distribution of
face-on central disk surface brightness peaks at \tsim22.5 mag\persecsq,
has a mean of 22.0 mag\persecsq, and a FWHM of 2.6 mag\persecsq. 

The sample contains several examples of low surface brightness (LSB)
galaxies and galaxies containing LSB disks, the most extreme of which
are A01374\-$+$1539B, A12263\-$+$4331 and A11238\-$+$5401 (but see also
A10114\-$+$0716 and IC~673 for examples of LSBs with bulges).  Although
this sample, as most magnitude limited samples, is deficient in LSBs
(Sprayberry \etal 1995; McGaugh, Schombert \& Bothun 1995) {\em we do
not omit this part of the galaxy population}. 

We do not have any examples of the very brightest end of the cluster
luminosity function, represented by the brightest cluster galaxies, in
our final sample.  These systems are very rare, however, in terms of
their number frequency averaged over the entire field, and in terms of
luminosity density.  We also lack examples of galaxies at the very
faintest end of the luminosity function (at $\MB>-14$), due to the
negligible sample volume of the CfA~I catalog at these faint absolute
magnitudes.  With few caveats, our sample is a faithful representation
of the local field galaxy population.

%%%%%%%%%%%%%%%%%%%%%%%%%%%%%%%%%%%%%%%%%%%%%%%%%%%%%%%%%%%%%%%%%%%%%%%%
% SECTION  SECTION  SECTION  SECTION  SECTION  SECTION  SECTION  SECTION
%%%%%%%%%%%%%%%%%%%%%%%%%%%%%%%%%%%%%%%%%%%%%%%%%%%%%%%%%%%%%%%%%%%%%%%%

\section{Observations, reduction and calibration}
\label{P-obsredcal}

\subsection{Observations}
\label{P-observations}

The photometric observations reported here were made with CCD cameras at
the F.L.~Whipple Observatory's 1.2 m telescope\footnote{The F.L.~Whipple
Observatory is operated by the Smithsonian Astrophysical Observatory,
and is located on Mt.~Hopkins in Arizona.}.  The data were obtained
during 50 dark nights between 1994 March and 1997 March.  We used both
front-side and thinned back-side illuminated versions of Loral CCDs with
2048$\times$2048, 15 \micron\ pixels.  This yields a scale of \tsim
0.32\arcsec\ per pixel, although we usually binned the images 2 by 2. 
The images cover a region \tsim 11\arcmin\ square.  For both CCDs the
readout noise was \tsim 8 e$^{-}$ RMS. 

We obtained $B$ and $R$ images with the front-side illuminated CCD,
typically integrating $2\times900$ seconds in $B$ and $2\times450$
seconds in $R$.  Beginning in 1995 October, the back-side illuminated
CCD camera was commissioned, allowing $U$ exposures.  We typically
integrated 900 seconds in $U$, $2\times450$ seconds in $B$ and
$2\times180$ seconds in $R$ with this CCD. 

The $U$ and $B$ filters closely follow the Johnson (1955) system, while
the $R$ filter follows the Cousins (1976) system.  The $B$ filter has a
small leak at 5600\AA\ that contributes \tsim 0.5\% to the total light
for a flat spectrum source.  On photometric nights, we observed several
Landolt (1992) standard star fields at low and high airmass for
photometric calibration.  We obtained short and long exposures to allow
the use of standard stars at a wide range of magnitudes and to check the
CCD linearity.  At least one standard star field was followed throughout
most of the night.  On photometric nights, we typically obtained short
galaxy exposures to calibrate the non-photometric observations.  The
diameters of stellar images on our frames are typically 1.5\arcsec\ to
2.0\arcsec\ FWHM.

\subsection{Data reduction}
\label{P-reduction}

The data were reduced within the IRAF environment using a custom
reduction pipeline that follows standard reduction techniques.  The
sequence of steps is: (1) interpolation over bad columns, dead and hot
pixels, (2) bias subtraction using the median of the bias frames,
followed by removal of the mean pixel value in the overscan region, (3)
dark current subtraction using a median of the dark exposures, (4)
subtraction of scattered light from a field flattening lens (October
1995 data only), and (5) flat fielding, correcting for the
pixel-to-pixel variations with dome flats and for the illumination
variations with twilight flats. 

Prior to flat fielding, the consistency of the flats from night to night
over each run was checked.  The field flattener/dewar window was cleaned
just before the observing run and then left undisturbed.  The flat
fields of separate nights correlate very well (frame to frame deviations
less than 0.2\%), allowing us to construct a single flat field for each
filter for an entire run. 

Following flat fielding, the images without bright stars are flat to
better than 0.8\% overall, improving to better than 0.5\% in the cropped
central 5.3\arcmin\ square portion of the CCD used for photometric
analysis of all but the most extended galaxies.  Even in the cropped
images the field of view is large enough for accurate determination of
the sky level. 

Ellipses were fit to all galaxy images to obtain radial intensity
profiles using the procedures described in Franx \etal (1989) and J\o
rgensen \etal (1992), and references therein.  Regions containing CCD
defects, cosmic ray events, stars, and background or companion galaxies
were excluded from the fit.  The width of the fitted elliptical annuli
grows by a factor 1.1 in each radial step.  This maintains a nearly
constant signal-to-noise-ratio in the azimutally averaged intensities by
compensating for the decline in surface brightness with radius.  The
fits were continued to the radius where only 60\% of the elliptical
annulus falls inside the image. 

We first fit the position of the galaxy center, fixing the position
angle and ellipticity at the UGC values (Nilson 1973), listed in
table~1.  Then, with the center fixed to the fitted position, the
ellipticity and position angles were fitted.  In each case where the fit
converged, the ellipticity and position angle in the outer parts of the
galaxies matched the UGC values within the errors (see also de~Jong and
van~der~Kruit (1994)), independent of the filter used.  We therefore
adopted the UGC ellipticities and position angles with two exceptions:
A12446\-$+$5155 has a position angle of 75 rather than 5\arcdeg, and
A11336\-$+$5829 has a position angle of 171 (or --9) rather than
9\arcdeg.  If the UGC omits a position angle for a round, very
disturbed, or very compact system, we use a position angle of 90\arcdeg. 
The final radial intensity profile was then fitted with the galaxy
center, ellipticity, and position angle fixed at all radii. 

Our goal is to obtain accurate photometry and internally consistent
colors and surface brightness profiles, not to investigate the details
of the light distribution.  Ellipses will fail to adequately represent
the isophote shapes of some of the galaxies in the sample, including
barred galaxies and many late type galaxies.  In addition, many late
type galaxies have ellipticities and position angles that vary with
radius, and in some cases the isophote centers wander. 

The non-photometric data were calibrated with the help of the short
calibration exposures obtained during photometric conditions.  In a
few cases, no photometric image was available, and we averaged the
photometric scaling for the bracketing images if both agreed to better
than \tsim 30\%. These exceptions have been clearly marked in the
figures and data tables presenting our surface brightness profiles and
measurements (section~\ref{P-results}).

The sky value was determined from the mean of the intensities of the 5
outermost annuli.  These annuli extend to the edge of the images and
contain many thousands of pixels.  The variance in the sky level is
dominated by systematic errors, which we estimate from the peak-to-peak
variation in the background level over the images, excluding linear
gradients.  The systematic errors in the sky level are typically 0.5\%,
except where bright neighbouring galaxies or stars are present.

\subsection{Calibration}
\label{P-calibration}

%%%%%%%%%%%%%%%%%%%% PLACE TABLE IN THIS SUBSECTION %%%%%%%%%%%%%%%%%%%%
%
%\placetable{P-Tab.photcal}
%\begin{table}\dummytable\label{P-Tab.photcal}\end{table}
%
%%%%%%%%%%%%%%%%%%%%%%%%%%%%%%%%%%%%%%%%%%%%%%%%%%%%%%%%%%%%%%%%%%%%%%%%

For each flux standard star we measured instrumental magnitudes in a
14\arcsec\ diameter circular aperture for conformity with Landolt's
(1992) photo-electric flux measurements.  For each filter on each
photometric night we used an interactive fitting procedure to find the
photometric zeropoint, atmospheric extinction coefficient and color term
coefficient.  A night was declared non-photometric if: (1) we detected
clouds or brightness fluctuations of the guide star, or (2) the RMS
scatter in the photometric fit to the standard stars exceeded 0.035,
0.028, and 0.025 mag in $U$, $B$, and $R$ respectively, or (3) the
scatter in the photometric fit was time varying.  Table~2 gives typical
photometric constants for each filter and both CCD detectors.  The
sigmas in this table denote the spread (not the errors) of the fitted
coefficients in almost 2 years of standard star data for each CCD.

%%%%%%%%%%%%%%%%%%%%%%%%%%%%%%%%%%%%%%%%%%%%%%%%%%%%%%%%%%%%%%%%%%%%%%%%%
%         TABLE 2: AVERAGE PHOTOMETRIC CALIBRATION CONSTANTS            %
%%%%%%%%%%%%%%%%%%%%%%%%%%%%%%%%%%%%%%%%%%%%%%%%%%%%%%%%%%%%%%%%%%%%%%%%%
\vbox{
\begin{center}
\setlength{\tabcolsep}{3pt}
\centerline{\sc Table 2}
\centerline{\footnotesize\sc Average photometric constants 
            FLWO 1994--1997\label{P-Tab.photcal}}
\vspace*{0.2cm}
\scriptsize
\begin{tabular}{llccccccc}
\hline
\hline\noalign{\vskip 0.1cm}
filter & CCD & \multicolumn{2}{c}{zeropoint} & 
\multicolumn{2}{c}{extinction coeff.} &
\multicolumn{3}{c}{color term coeff.} \\
       &     &    $pz$    &   $\sigma_{pz}$  &
 $pe$  &   $\sigma_{pe}$  &
 $pc$  &   $\sigma_{pc}$  & color \\\noalign{\vskip 0.1cm}
\hline\noalign{\vskip 0.1cm}
%--------------------------------------------------------------------------
% filt   chip    pzero  epzero   pext   epext     pcol     epcol   color
%--------------------------------------------------------------------------
 $U$   & thick & 19.97 & 0.08 & 0.482 & 0.039 & $-$0.049 & 0.007 & $U-R$ \\
 $U$   & thin  & 21.76 & 0.06 & 0.482 & 0.039 & $-$0.046 & 0.006 & $U-R$ \\
 $B$   & thick & 21.61 & 0.09 & 0.252 & 0.022 & $-$0.067 & 0.010 & $B-R$ \\
 $B$   & thin  & 23.24 & 0.05 & 0.252 & 0.022 & $-$0.024 & 0.008 & $B-R$ \\
 $V^a$ & thick & 21.93 & ---  & 0.112 & ---   & $+$0.008 & ---   & $B-V$ \\
 $V^a$ & thin  & 23.04 & 0.08 & 0.151 & 0.034 & $+$0.043 & 0.015 & $B-V$ \\
 $R$   & thick & 22.10 & 0.11 & 0.115 & 0.020 & $+$0.035 & 0.005 & $B-R$ \\
 $R$   & thin  & 23.14 & 0.06 & 0.115 & 0.020 & $+$0.027 & 0.007 & $B-R$ \\
 $I^a$ & thin  & 22.92 & 0.05 & 0.053 & 0.023 & $-$0.085 & 0.012 & $R-I$ \\
\noalign{\vskip 0.02cm}
%---------------------------------------------------------------------------
\hline
\end{tabular}
\end{center}\par
\parbox[b]{0.45\txw}{\footnotesize $^a$ $V$ and $I$ filter standards were 
observed for co-scheduled monitoring programs.}
}\vspace*{0.3cm}

\noindent The radial intensity profiles were then photometrically
calibrated, solving simultaneously the calibration equations for $B$ and
$R$, or $U$ and $R$, using $(B-R)$ and \UR\ color terms, respectively. 
The details of this procedure are found in appendix~\ref{P-App.calib}. 

Finally, the calibrated surface brightness profiles were combined if
multiple exposures were available.  Because the repeated exposures were
frequently obtained on different nights and even on different observing
runs, the scatter in the individual profiles allows us to estimate the
photometric errors.  Where the variance of the calibrated profiles
exceeds the formal photometric errors (see appendix~\ref{P-App.calib}),
we adopt this larger variance as our photometric error.  In most cases
this applies only to the inner few arcseconds of the profiles, where
seeing changes (1.2--2.5\arcsec) cause up to \tsim 0.2 mag\persecsq\
variations between individual profiles.  We conclude that the shapes of
the derived surface brightness profiles are reliable. 

The radial surface brightness profiles were corrected for Galactic
exinction, estimated from the \HI maps of Burstein \& Heiles (1984) and
using extinction ratios $A_R/A_B=0.588$ and $A_U/A_B=1.156$ (Rieke \&
Lebofsky 1985) to derive the corrections in $R$ and $U$.  No corrections
for cosmological effects (dimming and $k$-correction) were applied. 

The profiles were used to measure isophotal and fractional magnitudes,
and corresponding radii and colors (section~\ref{P-radmags}).

For our photometric analysis it sufficed to match the individual surface
brightness profiles of a galaxy.  The individual images were not shifted
to a common center, nor rebinned to a common position angle and pixel
scale.  Stars were not removed from the images.  Also, we did not remove
residual linear gradients in the background as they dropped out in the
fitting of the ellipses.  Although, in principle, these data are
suitable for two-dimensional bulge/bar/disk de-convolutions and other
detailed studies of the light distribution, at present the data are not
in an easily distributable format.

\subsection{Total magnitudes: extrapolation to infinity}
\label{P-totmags}

Following Han~(1992) we extrapolated the radial surface brightness
profiles to infinity to obtain total magnitudes, under the assumption
that the light at large radii can be approximated by an exponential
disk.  Exponentials were fit to the outer parts of each radial surface
profile over a 2.172 magnitude interval ending 0.3 mag\persecsq\ above
the $1\,\sigma$ limiting isophote.  For a purely exponential profile
this surface brightness interval corresponds to exactly two exponential
scale lengths.  Our profiles are sufficiently deep that the details of
the light fall off do not seriously affect the result.  We might
underestimate the luminosity of an $R^{1/4}$ profile elliptical by \tsim
0.1 mag with this procedure.  For disk galaxies the extrapolation
procedure should work very well unless the disk is truncated near the
outer fit interval.  However, we found no convincing evidence for disk
truncation in our sample.  Furthermore, the galaxies requiring the
largest corrections from isophotal to total magnitude follow an
exponential profile to the faintest regions sampled in our data.  The
average limiting isophotes for our data are \muU=26.7, \muB=27.2, and
\muR=26.1, with a spread of \tsim 0.7 mag\persecsq. 

Absolute magnitudes were calculated using the total magnitudes and the
galaxy recessional velocities with respect to the centroid of the
Local Group, correcting for Virgo infall.  We used the infall model of
Kraan-Korteweg \etal (1984) which has the Local Group falling in Virgo
at 220 \kms; no other corrections for peculiar motion were applied.

\subsection{Evaluation of the data quality}
\label{P-evaluation}

\subsubsection*{\footnotesize Internal consistency and error analysis}
\label{P-internalQ}

For most galaxies, more than one long exposure per filter is available. 
To evaluate the internal accuracy of our photometry we compare total and
isophotal magnitudes measured in the radial surface brightness profiles
derived from the individual exposures. In the many cases where we
compare data from before October 1995 with later observations, we
compare measurements obtained with different cameras.

We find RMS differences between individual measurements of the total
magnitude (expressed in magnitudes and in units of the error of
measurement) of 0.05 mag ($0.86\sigma_U$), 0.03 mag ($0.72\sigma_B$) and
0.03 mag ($0.66\sigma_R$) in $U,B$ and $R$, respectively.  For the
majority of the sample, however, individual measurements agree to better
than the values quoted above; typical values would be 0.030, 0.013, and
0.012 mag in $U,B$ and $R$.  This result is consistent with the
estimated accuracy based on the errors in the fitted zeropoints\vfill

%%%%%%%%%%%%%%%%%%%%%%%%%%%%%%%%%%%%%%%%%%%%%%%%%%%%%%%%%%%%%%%%%%%%%%%%%
%      FIGURE 2: COMPARISON WITH SURFACE PHOTOMETRY OTHER AUTHORS	%
%%%%%%%%%%%%%%%%%%%%%%%%%%%%%%%%%%%%%%%%%%%%%%%%%%%%%%%%%%%%%%%%%%%%%%%%%
\noindent\leavevmode
\framebox[\txw]{
   \centerline{\parbox[c]{0.70\txw}{
       \centerline{\rule[-0.60\txw]{0pt}{1.20\txw}
       \centerline{\parbox[c]{0.7\txw}{\Large\it
          \hfill``fig2.jpg''\hfill\null\\$\quad$\\
          A 600 dpi postscript version of this preprint, including all
	  figures and the atlas of galaxy images and radial profiles, can
          be retrieved from URL\  http://www.astro.rug.nl/\tsim nfgs/}
       }}
   }}
}\par\vspace*{3mm}

\noindent\makebox[\txw]{
\centerline{
\parbox[t]{\txw}{\footnotesize {\sc Fig.~2 ---} Comparison of our
surface brightness profiles and profiles published by other authors. 
The difference $\Delta\mu $ is in the sense (our profiles -- literature
profiles).  Offsets in photometric zeropoint are indicated where
applied.  As an aid in distinguishing significant differences between
profiles we plot $1\,\sigma$ deviations as thin vertical bars.  Key to
the abbreviations used: SB'99=Swaters and Balcells (1999),
TVPHW'96=Tully \etal (1996), dJvdK'94=de~Jong and van der Kruit (1994),
PDDIC'90=Peletier \etal (1990), and vdK'87=van der Kruit (1987).  } }
}\vfill\newpage
%
%%%%%%%%%%%%%%%%%%%%%%%%%%%%%%%%%%%%%%%%%%%%%%%%%%%%%%%%%%%%%%%%%%%%%%%%

\noindent and in the photometric scaling of non-photometric exposures. 
The measured differences do not depend on galaxy color, morphological
type, date of observation, absolute magnitude, nor on effective surface
brightness. 

A more demanding test of our photometry, however, is a comparison with
surface photometry published by other authors.

\subsubsection*{\footnotesize Comparison with other authors}
\label{P-externalQ}

Our sample has 16 galaxies in common with the dwarf galaxy sample of
Swaters \& Balcells (in prep.; hereafter SB99), most of which were
observed by these authors in $R$ only; they observed only two of the 16
in $B$.  Rob Swaters kindly provided us with his surface brightness
profiles, allowing us to evaluate the external accuracy of our
photometry.  The results are presented in figure~2.  We find good
agreement for all galaxies that were observed under photometric
conditions in both studies.  Where SB99 note that cirrus was present
(Swaters 1998, private comm.), we are able to compare the shape of the
two sets of profiles after applying an offset (see figure text).  Where
necessary, we correct the SB99 results to account for a different
assumed ellipticity; this correction never exceeded \tsim 10\%. 

We have 6 galaxies in common with another recent study of galaxies in
the Ursa Major cluster (Tully \etal 1996; hereafter TVPHW96).  These
authors present $B,R,I$ and $K$ data.  Marc Verheijen kindly provided us
with tables of the data plotted in TVPHW96.  Here again the photometric
zeropoints agree well.  In the worst cases, the surface brightness
profiles at very large radii deviate by a few tenths of a magnitude per
square arcsecond.  For A11547\-$+$4933 our photometry is not very deep,
and the $R$ profiles of SB99 and TVPHW96 seem to be of higher quality. 

We also show comparions for a small number of galaxies in common with
Peletier \etal (1990, PDDIC90), de~Jong \& van~der~Kruit (1994,
dJvdK94), and the earlier photographic work of van~der~Kruit (1987). 
Comparisons with Kent (1988) for NGC~2844 and Patterson \& Thuan (1996,
PT96) for A12263\-$+$4331, are not shown in figure~2, but are poor.  We
note that our $R$ profile of A12263\-$+$4331 agrees well with SB99.

%%%%%%%%%%%%%%%%%%%%%%%%% PLACE FIGURES HERE %%%%%%%%%%%%%%%%%%%%%%%%%%%
%
\placefigure{P-Fig.littcomp}
%
%%%%%%%%%%%%%%%%%%%%%%%%%%%%%%%%%%%%%%%%%%%%%%%%%%%%%%%%%%%%%%%%%%%%%%%%

%%%%%%%%%%%%%%%%%%%%%%%%%%%%%%%%%%%%%%%%%%%%%%%%%%%%%%%%%%%%%%%%%%%%%%%%
% SECTION  SECTION  SECTION  SECTION  SECTION  SECTION  SECTION  SECTION
%%%%%%%%%%%%%%%%%%%%%%%%%%%%%%%%%%%%%%%%%%%%%%%%%%%%%%%%%%%%%%%%%%%%%%%%

\section{Photometric results}
\label{P-results}

\subsection{Data presentation}
\label{P-atlas}
\label{P-radmags}

Here we describe our primary data products, galaxy images and radial
profiles of surface brightness and color, and present our photometric
measurements.  For 198 galaxies, our atlas includes logarithmically
scaled renditions of the $B$ images, $B$ surface brightness profiles,
and plots of \UB\ and \BR\ colors as a function of radius (figure~3). 
The galaxies have been sorted by morphological type, and within each
type by blue luminosity. 

The radial profiles are not corrected for seeing, and are not a faithful
representation of the surface brightness at small radii less than a few
arcseconds.  In some cases the inner part of a surface brightness
profile appears to flatten over a larger region than that affected by
seeing.  This extended flattening results from a bar, irregularly
distributed central star formation, dust, or multiple nuclei.  Notes on
individual galaxies and their radial profiles are collected in
Appendix~\ref{P-App.objects}.

%%%%%%%%%%%%%%%%%%%%%%%%%%% PLACE ATLAS HERE %%%%%%%%%%%%%%%%%%%%%%%%%%%
%
\placefigure{P-Fig.atlas}
%
%%%%%%%%%%%%%%%%%%%%%%%%%%%%%%%%%%%%%%%%%%%%%%%%%%%%%%%%%%%%%%%%%%%%%%%%

The photometric measurements are collected in table~3.  Columns (1)
through (3) list the galaxy identifications.  Column (4) gives the
distances in Mpc derived from the Virgocentric flow corrected velocities
with respect to the centroid of the Local Group (column (12) of
table~1), assuming \Ho=100 \kmsMpc.  Columns (5) and (6) list the $B$
apparent and absolute total magnitudes.  In columns (7) and (8) we give
the elliptical radii, $r^B_{26}$, at which the surface brightness has
dropped to \muB=26 mag\persecsq\ and the corresponding isophotal
magnitude, $B_{26}$.  Columns (9) and (10) list the effective (\ie
half-light) elliptical radius in $B$, \rBe, and the surface brightness,
\muBe, at that radius.  Columns (11) and (12) contain the effective \UB\
and \BR\ colors as measured within the effective radii in $B$.  The
final two columns list the color differences, $\Delta (U-B)_{25-75}$ and
$\Delta (B-R)_{25-75}$, between the inner and outer parts of the
galaxies, as defined below.

%%%%%%%%%%%%%%%%%%%%%%%%%%% PLACE TABLE HERE %%%%%%%%%%%%%%%%%%%%%%%%%%%
%
%\placetable{P-Tab.photmeas}
%\begin{table}\dummytable\label{P-Tab.photmeas}\end{table}
%
%%%%%%%%%%%%%%%%%%%%%%%%%%%%%%%%%%%%%%%%%%%%%%%%%%%%%%%%%%%%%%%%%%%%%%%%

Several authors have investigated the Zwicky magnitude scale.  Bothun \&
Cornell (1990) showed for a sample of 107 spiral galaxies with
$m_Z\geq14.0$ that, although Zwicky magnitudes are not isophotal
magnitudes, the scatter around isophotal magnitudes varying from
$B_{22}$ to $B_{27}$ is of the order of 0.3 mag.  In the mean, $m_Z$
corresponds to $B_{26}$ with a scatter of 0.31 mag.  In figure~4{\em a}
we plot the isophotal magnitudes measured within the elliptical aperture
corresponding to the \muB=26 mag\persecsq\ isophote versus the
photographic Zwicky magnitudes.  We find for our sample a similar
correspondence between $m_Z$ and $B_{26}$ with observed RMSs of 0.30 mag
for galaxies with $m_Z<14$ and of 0.42 mag for $m_Z\geq 14$.  Using
total magnitudes instead of $B_{26}$ (figure~4{\em b}) gives a nearly
identical result.  Figure~4{\em c} shows that there is a relation
between the difference $(m_Z-B_{26})$ and $B_{26}$ for $m_Z>14$: Zwicky
tends to overestimate the magnitude (\ie underestimate the brightness)
of the brighter galaxies and underestimate the magnitudes of the fainter
ones.  The dependence of the magnitude difference on galaxy type is
weak, with a very large \vfill

%%%%%%%%%%%%%%%%%%%%%%%%%%%%%%%%%%%%%%%%%%%%%%%%%%%%%%%%%%%%%%%%%%%%%%%%%
%      FIGURE 3: (ATLAS) B-FILTER IMAGERY AND PHOTOMETRIC PROFILES	%
%%%%%%%%%%%%%%%%%%%%%%%%%%%%%%%%%%%%%%%%%%%%%%%%%%%%%%%%%%%%%%%%%%%%%%%%%
\renewcommand{\textfraction}{0.0000}
\noindent\leavevmode\framebox[\txw]{
   \centerline{\parbox[c]{\txw}{
       \centerline{\rule[-0.55\txw]{0pt}{1.10\txw}
       \parbox[c]{0.7\txw}{\Large\it
	  Atlas of images and radial profiles --- pages 12 through 46\\
          (``fig3\_01.jpg'' through ``fig3\_35.jpg'')\\$\quad$\\
          A 600 dpi postscript version of this preprint, including all
	  figures and the atlas of galaxy images and radial profiles, can
          be retrieved from URL\  http://www.astro.rug.nl/\tsim nfgs/}}}
   }
}\par\vspace*{3mm}\noindent\leavevmode\makebox[\txw]{\centerline{
\parbox[t]{\txw}{\footnotesize {\sc Fig.~3 ---} The atlas of images
and radial profiles.  Greyscale renditions of a $B$ filter image and the
radial dependence of the surface brightness in $U,B,R$ (top panel) and
corresponding colors (bottom panel) are presented for each of the
observed galaxies.  The galaxies have been ordered according to their
morphological type and per type according to their absolute $B$
magnitude (both given in the images).  North is up and East is to the
left.  The original CCD images were much larger than the portion
centered on the galaxies presented here.  The image scale is given by a
scale bar of the indicated number of kpc (assuming H$_0$=100 km s$^{-1}$
Mpc$^{-1}$), and is given on the top axis of the radial profile plots
for reference.  On the ordinate of the radial profile plots $r$ denotes
the elliptical radius in arcseconds.  The extrapolation of the surface
brightness profiles beyond the limiting isophote (see text) is indicated
by grey lines.  The effective radius in $B$ is indicated by a small
vertical arrow.  Shallow profiles (only short photometric exposures are
available) are indicated by (s), truncated profiles (due to overlapping
neighbours or bright stars close to a galaxy) are indicated by (t), and
profiles with uncertain photometric zeropoints by (z) or (z!).  The
names of the two galaxies not part of the statistical sample are placed
in parentheses.\\
Compact ellipticals (page 12) through Irregulars (page 46).}
} }\newpage

\setcounter{page}{47}
%
%%%%%%%%%%%%%%%%%%%%%%%%%%%%%%%%%%%%%%%%%%%%%%%%%%%%%%%%%%%%%%%%%%%%%%%%

%%%%%%%%%%%%%%%%%%%%%%%%%%%%%%%%%%%%%%%%%%%%%%%%%%%%%%%%%%%%%%%%%%%%%%%%%
%    FIGURE 4: COMPARISON OF CCD B AND PHOTOGRAPHIC ZWICKY MAGNITUDES	%
%%%%%%%%%%%%%%%%%%%%%%%%%%%%%%%%%%%%%%%%%%%%%%%%%%%%%%%%%%%%%%%%%%%%%%%%%
\noindent\leavevmode
\makebox[\txw]{
   \centerline{
      \epsfig{file=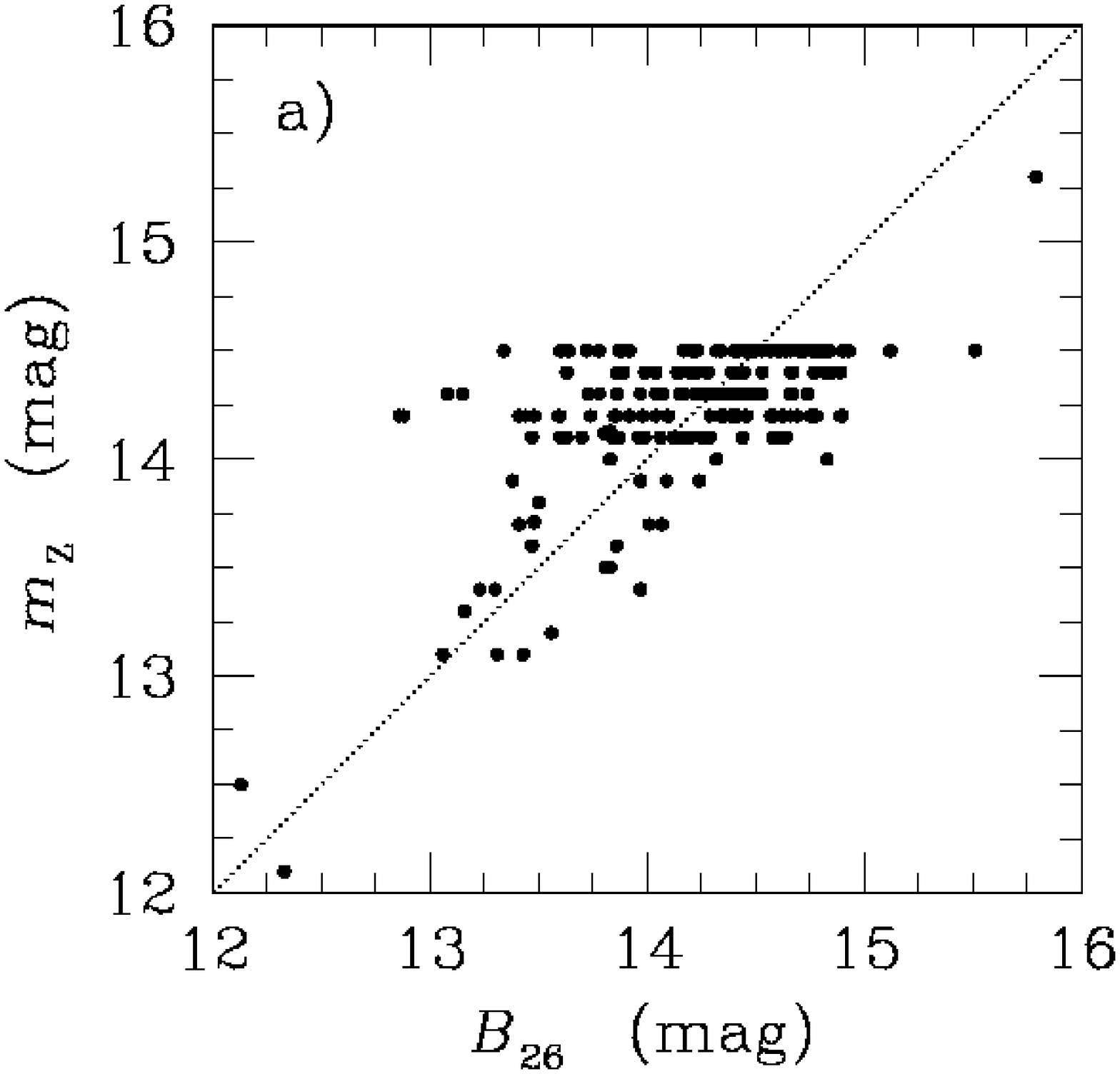,width=0.32\txw,clip=}\hspace*{0.05\txw}
      \epsfig{file=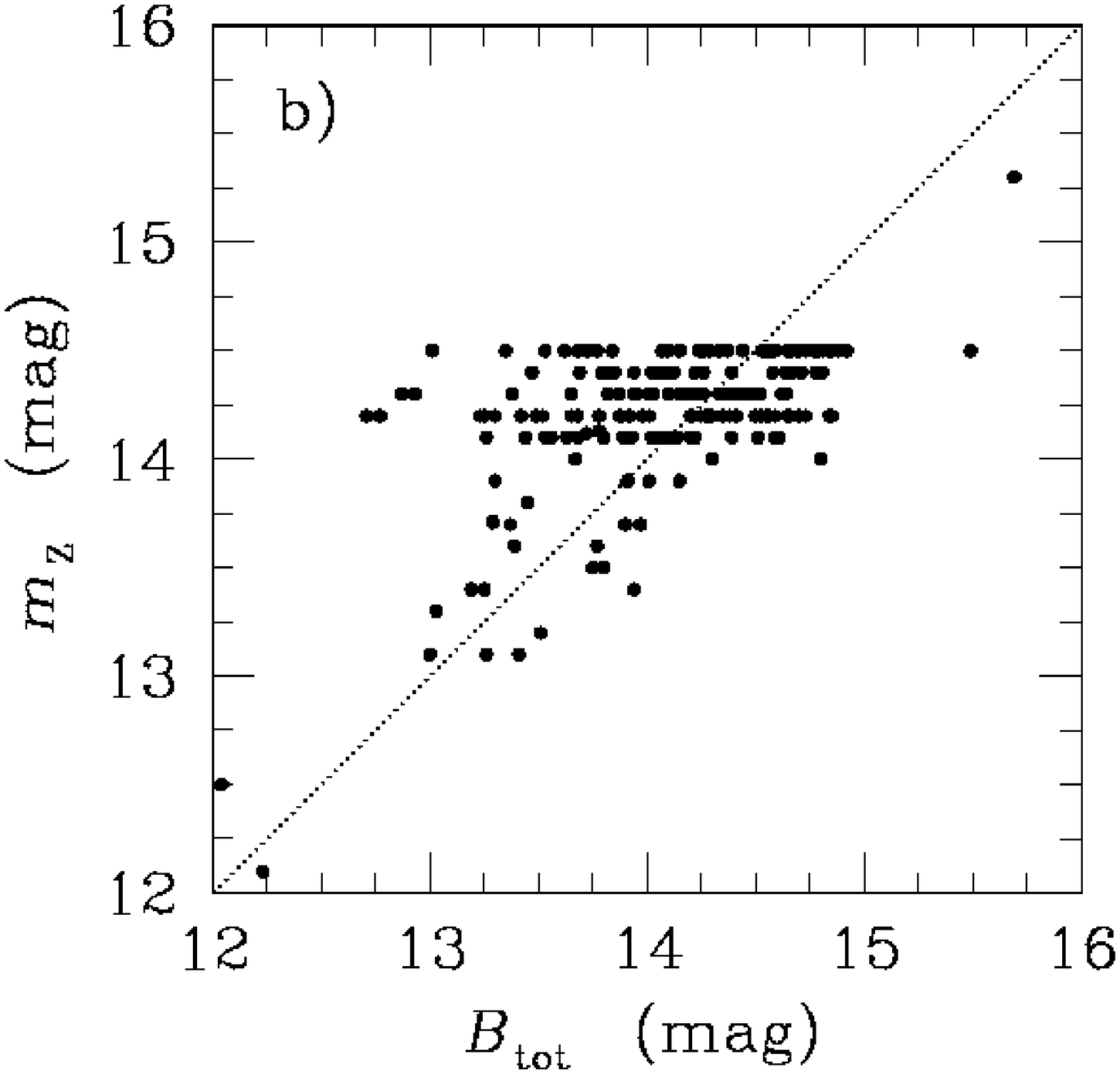,width=0.32\txw,clip=}
   }
}\par\vspace*{0.05\txw}\noindent\leavevmode
\makebox[\txw]{
   \centerline{
      \epsfig{file=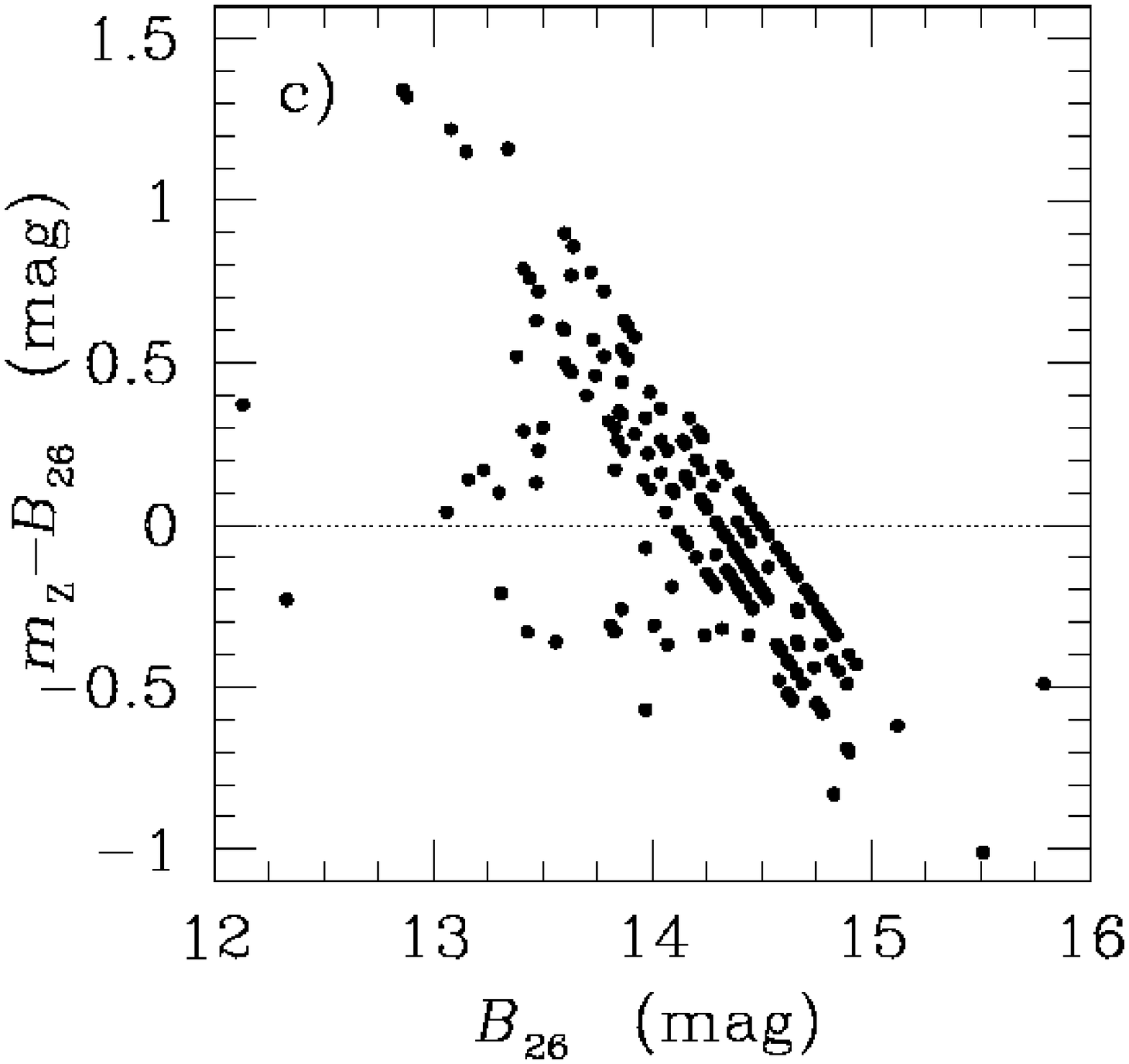,width=0.32\txw,clip=}\hspace*{0.05\txw}
      \epsfig{file=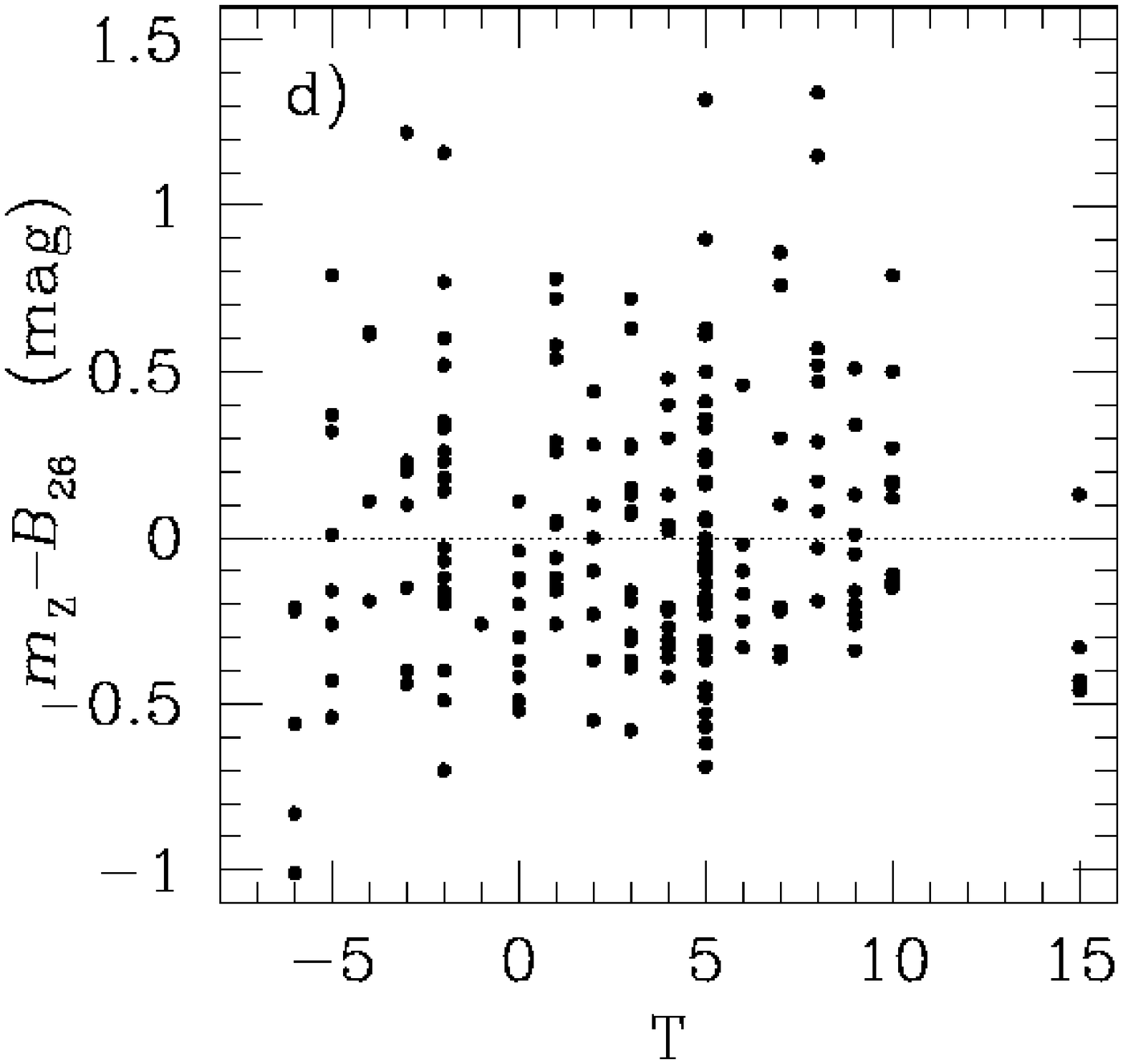,width=0.32\txw,clip=}
   }
}\par\noindent\leavevmode
\makebox[\txw]{
\centerline{
\parbox[t]{\txw}{\footnotesize {\sc Fig.~4 ---} Comparison of
photographic Zwicky magnitudes and a) $B_{26}$ isophotal magnitudes, and
b) total $B$ magnitudes.  For $m_Z$ brighter than 14 the observed RMS is
0.30 in both cases; for $m_Z\geq 14$ the RMS is 0.42 and 0.45 mag for
isophotal and total magnitudes, respectively.  c) Difference
$m_Z-B_{26}$ as a function of $B_{26}$.  For $m_Z>14$ this difference is
correlated with isophotal magnitude.  d) $m_Z-B_{26}$ as a function of
galaxy type.  A weak trend of larger differences towards later galaxy
type may be discerned, but the scatter on this trend is large.  } }
}\vspace*{0.5cm}

%
%%%%%%%%%%%%%%%%%%%%%%%%%%%%%%%%%%%%%%%%%%%%%%%%%%%%%%%%%%%%%%%%%%%%%%%%

\noindent scatter: Zwicky magnitudes for late type galaxies tend to be
somewhat overestimated, while those for early type galaxies are
underestimated.  This dependence on morphological type was also inferred
by Bothun \& Cornell (1990) based on weak trends of $(m_Z-B_{26})$ with
a light concentration parameter and average surface brightness in their
sample.

\subsection{Properties of our sample}

In Figure~5{\em a} we plot the distribution of total absolute $B$
magnitudes as a function of numeric galaxy type.  We show all galaxies,
including active galaxies (starbursts or active galactic nuclei (AGN)). 
The range in luminosity and type is large (\cf figure~1), and galaxies
of each type are distributed over a large range in absolute magnitude. 
However, the lack of intrinsically bright ($\MB\lesssim -19$) late type
spirals and irregular systems ($\mbox{T}>6$), and of low luminosity
early and intermediate type spiral galaxies ($1<\mbox{T}<5$) is evident. 
This corresponds to the type--magnitude relation for spiral galaxies
later than Sa found by other authors (see Sandage \& Binggeli 1984,
their figure~1).\vfill

\null\vspace*{0.77\txw}\noindent Our sample does include a number of
low luminosity early type S0 and elliptical systems.  The BL~Lac galaxy
Mrk~421 and the Seyfert~{\sc i} galaxies are amongst the intrinsically
brightest galaxies in our sample. 

In figure~5{\em b} we plot the distribution of the effective \BR\ color
(measured within the effective radius in $B$) as a function of galaxy
type.  The bulk of the galaxies become progressively bluer towards later
types.  This is not a new result, but the scatter on this trend is
surprisingly small ($\sigma \BRe=0.19$ mag).  Nonetheless, there are
examples of very blue early-type and reddish late-type galaxies.  The
next panel shows that the blue early type galaxies tend to be low
luminosity systems.  If redshifts are available, color may be used to
estimate a galaxy's broad type class (E,S,Irr) to a similar precision
(\tsim 5 T types) as measures of galaxy asymmetry and concentration of
the light (Abraham \etal 1996; see also Jansen \etal (in prep.) for an
application of that method to our sample).  This result is remarkable if
not entirely new (de Vaucouleurs 1977; Tinsley 1980). \vfill

In figure~5{\em c} the distribution of effective galaxy colors is given
as a function of absolute $B$ magnitude.  Luminous systems tend to be
redder than fainter ones.  This is a relationship well-established for
early type galaxies (\eg Sandage \& Visvanathan 1978; Peletier \etal
1990; Ellis \etal 1997).  The scatter in color for all galaxy types
combined is larger ($\sigma\BRe=0.23$ mag) than that for the color--type
relation.  The scatter for the separate type classes E--S0, S0/a--Sab,
Sb--Sc and Scd--Irr is 0.10, 0.15, 0.17 and 0.13 mag (median absolute
deviation (MAD)), respectively.  It is striking that ellipticals, S0's,
and early type spiral galaxies occupy a different region in the plot
than the late type spiral and irregular galaxies.  The intermediate type
spirals can be found mixed with both early and late type systems.  The
two distributions are offset from one another by \tsim 0.4 mag in \BRe\
color, but there is a large overlap.  This result persists if total
colors are used or colors measured in the region between the radii
containing 25\% and 75\% of the light in $B$. 

Figure~5{\em d} is the color--color diagram of effective \UB\ versus
\BR\ colors.  The bulk of the normal galaxies follow a relation with
only a small scatter (Tinsley 1980).  The ``abnormal'' galaxies
(harboring star bursts or an active nucleus, or interacting with another
galaxy) show a much larger scatter (Larson \& Tinsley 1978; Tinsley
1980).  Again, we see that late and early type galaxies tend to occupy
different regions in the plot, with early type galaxies located at
redder colors.  The relations followed by early and late type systems
differ in slope, with the early type galaxies following a steeper slope. 
The dust extinction vector points nearly along the relation followed by
the intermediate and late type spirals and irregular systems.  Starburst
galaxies tend to lie below and to the left of the main relation, \ie at
bluer \UB\ colors for a given \BR\ color. 

In the following plots, figures~5{\em e} and {\em f}, we will plot only
measurements for the 168 normal (non-AGN, non-starburst) galaxies in our
sample with high quality photometry.  Figure~5{\em e} shows the
(trivial) relation, that larger galaxies are more luminous.  We plot the
logarithm of the effective radius in kpc as a function of absolute
magnitude.  As more strongly concentrated systems have smaller effective
radii at a given absolute magnitude, we expect the early type galaxies
and spirals with significant bulges to follow a relation offset from
that followed by later type systems.  Indeed, we find that galaxies
fainter than $\MB\sim -19$ show a clear separation according to type. 
For the brightest systems, $\MB<-19$, this separation is not clear. 
Lower surface brightness galaxies lie above the mean relation for a
given absolute magnitude and given type. 

In figure~5{\em f} we plot the average $R$ surface brightness within the
effective radius, $\mu^R_{r\leq r^B_e}$, as a function of numeric galaxy
type.  The light distribution in early type galaxies is more
concentrated (\ie brighter in their inner parts) than that in later type
galaxies.  For spiral galaxies there is a strong trend towards fainter
$\mu^R_{r\leq r^B_e}$ at later types, consistent with a decreasing
contribution of bulges.  There is no clear trend within the late type
spirals and irregulars.  A similar conclusion is reached if the average
$B$ surface brightness within the effective radius is used, but the
trend is somewhat less pronounced.

%%%%%%%%%%%%%%%%%%%%%%%%%% PLACE FIGURE HERE %%%%%%%%%%%%%%%%%%%%%%%%%%%
%
\placefigure{P-Fig.photres}
%
%%%%%%%%%%%%%%%%%%%%%%%%%%%%%%%%%%%%%%%%%%%%%%%%%%%%%%%%%%%%%%%%%%%%%%%%

We now explore the radial color gradient in our sample galaxies.  We
measure \BR\ in the central region of the galaxy that contains 25\% of
the light, $\BR_{25}$, and in the region that contains the next 50\% of
the light, $\BR_{25-75}\;$.  We define $\Delta (B-R)_{25-75}$ as the
difference between these two measurements: $\BR_{25-75}-\BR_{25}\;$. 
$\Delta (U-B)_{25-75}$ is defined similarly.  If $\Delta (B-R)_{25-75}$
is negative, the inner parts of the galaxy are redder than the outer
parts. 

Figure~6{\em a} is a plot of $\Delta (B-R)_{25-75}$ against galaxy type
for the 168 ``normal'' galaxies.  The median color difference values for
each type have been connected by solid lines in this plot.  The color
difference between the inner and outer parts of a galaxy remains fairly
constant for most of the early type galaxies ($\mbox{T}\le 0$): the
inner parts are on average 0.09 mag (with a range of \tsim 0.15 mag)
redder than the outer parts.  Moving from S0/a to Sa, this difference
becomes larger.  The centers of spiral galaxies are redder by 0.18 mag,
on average (with a range of \tsim 0.30 mag), than their outer parts. 
The color differences decrease for late type spirals, reflecting the
decreasing prominence of bulges and the more extended star formation. 
The inner parts of irregulars can be either bluer or redder than the
outer parts.  We will return to this issue below.

%%%%%%%%%%%%%%%%%%%%%%%%%% PLACE FIGURE HERE %%%%%%%%%%%%%%%%%%%%%%%%%%%
%
\placefigure{P-Fig.coldiff}
%
%%%%%%%%%%%%%%%%%%%%%%%%%%%%%%%%%%%%%%%%%%%%%%%%%%%%%%%%%%%%%%%%%%%%%%%%

The bluing of galaxies at larger radii is not a new result (Sandage
1972; Persson, Frogel and Aaronson 1979; de~Jong 1996).  Recently,
however, Tully \etal (1996) describe an absolute magnitude dependent
behavior for galaxies in the Ursa Major cluster.  These authors find
that galaxies brighter than $\MB\sim -17$ become bluer with radius, and
that fainter galaxies show the opposite behavior.  We confirm this
result (figure~6{\em b}), although the transition from a bluing trend
with radius to a reddening trend may occur at a fainter limit.  We find
that {\em galaxies brighter than} $\MB=-17$ {\em are almost always
redder in their inner parts, while galaxies fainter than} $\MB\sim -17$
{\em are equally likely to redden or blue with radius.} Panel {\em c}
confirms that the majority of the galaxies that show \vfill

%%%%%%%%%%%%%%%%%%%%%%%%%%%%%%%%%%%%%%%%%%%%%%%%%%%%%%%%%%%%%%%%%%%%%%%%%
%     FIGURE 5: PHOTOMETRIC PROPERTIES OF THE NFGS SAMPLE GALAXIES	%
%%%%%%%%%%%%%%%%%%%%%%%%%%%%%%%%%%%%%%%%%%%%%%%%%%%%%%%%%%%%%%%%%%%%%%%%%
\noindent\leavevmode
\makebox[\txw]{
   \centerline{
      \epsfig{file=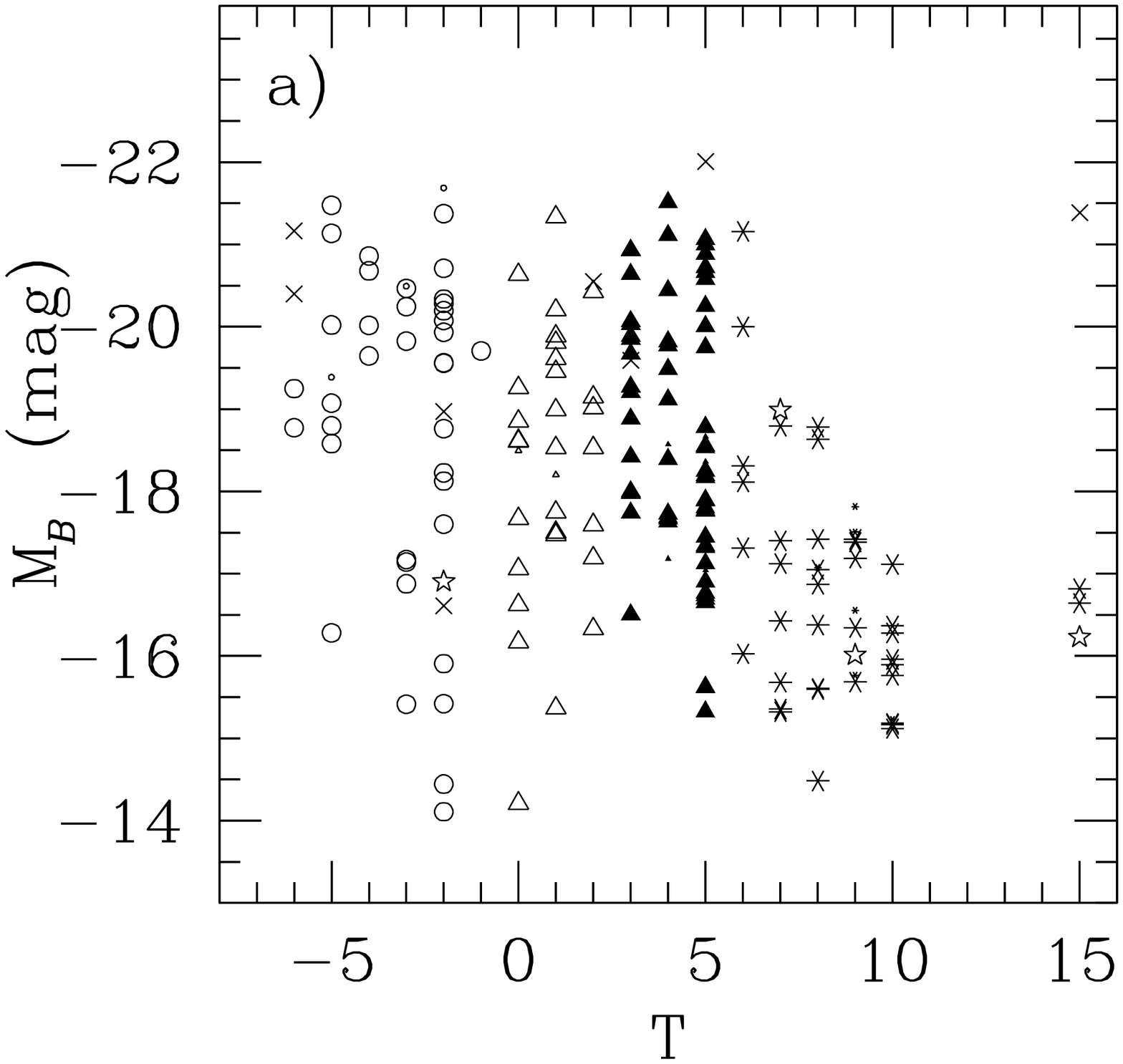,width=0.33\txw,clip=}
      \epsfig{file=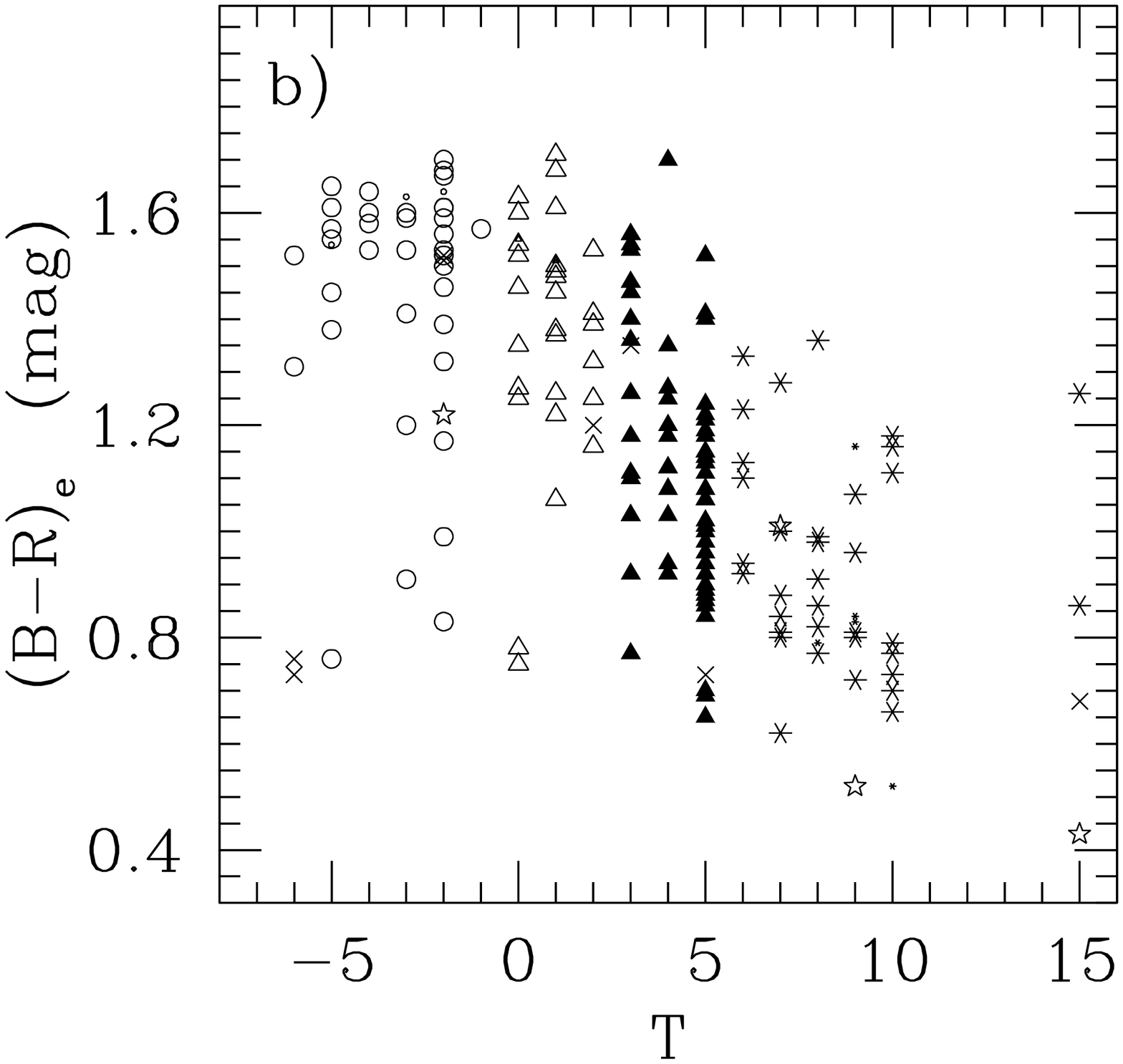,width=0.33\txw,clip=}
      \epsfig{file=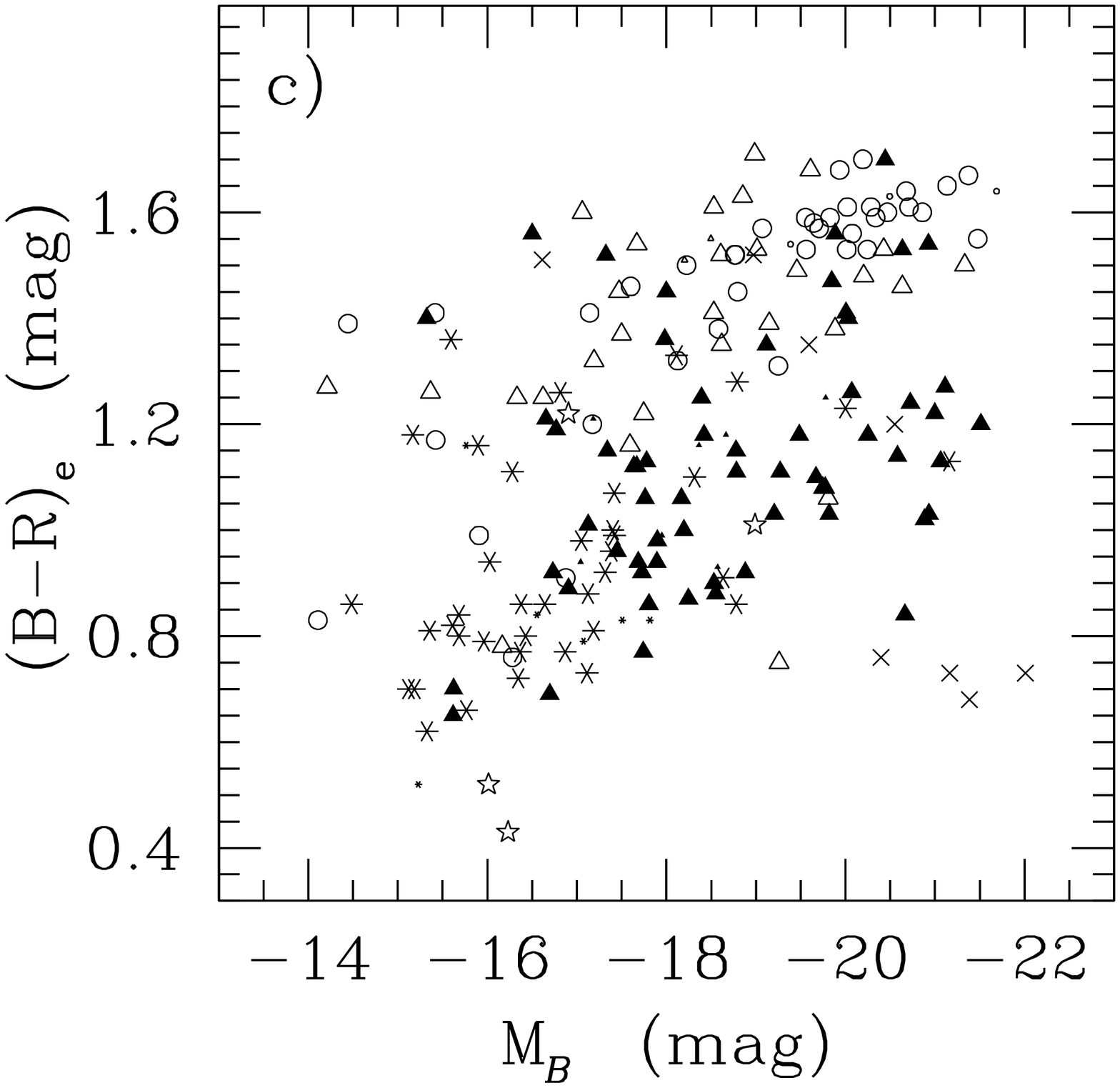,width=0.33\txw,clip=}
   }
}\par\noindent\leavevmode
\makebox[\txw]{
   \centerline{
      \epsfig{file=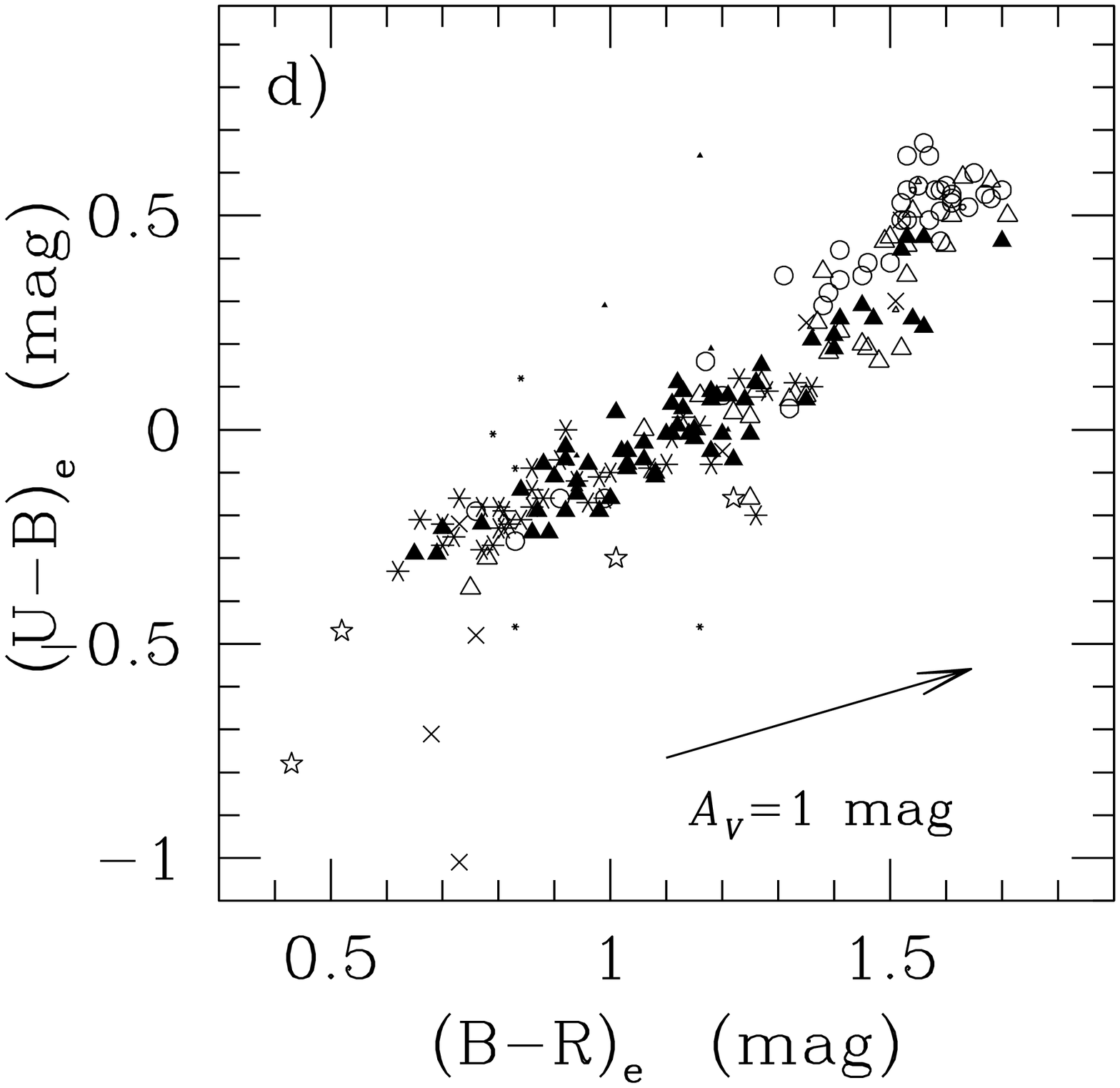,width=0.33\txw,clip=}
      \epsfig{file=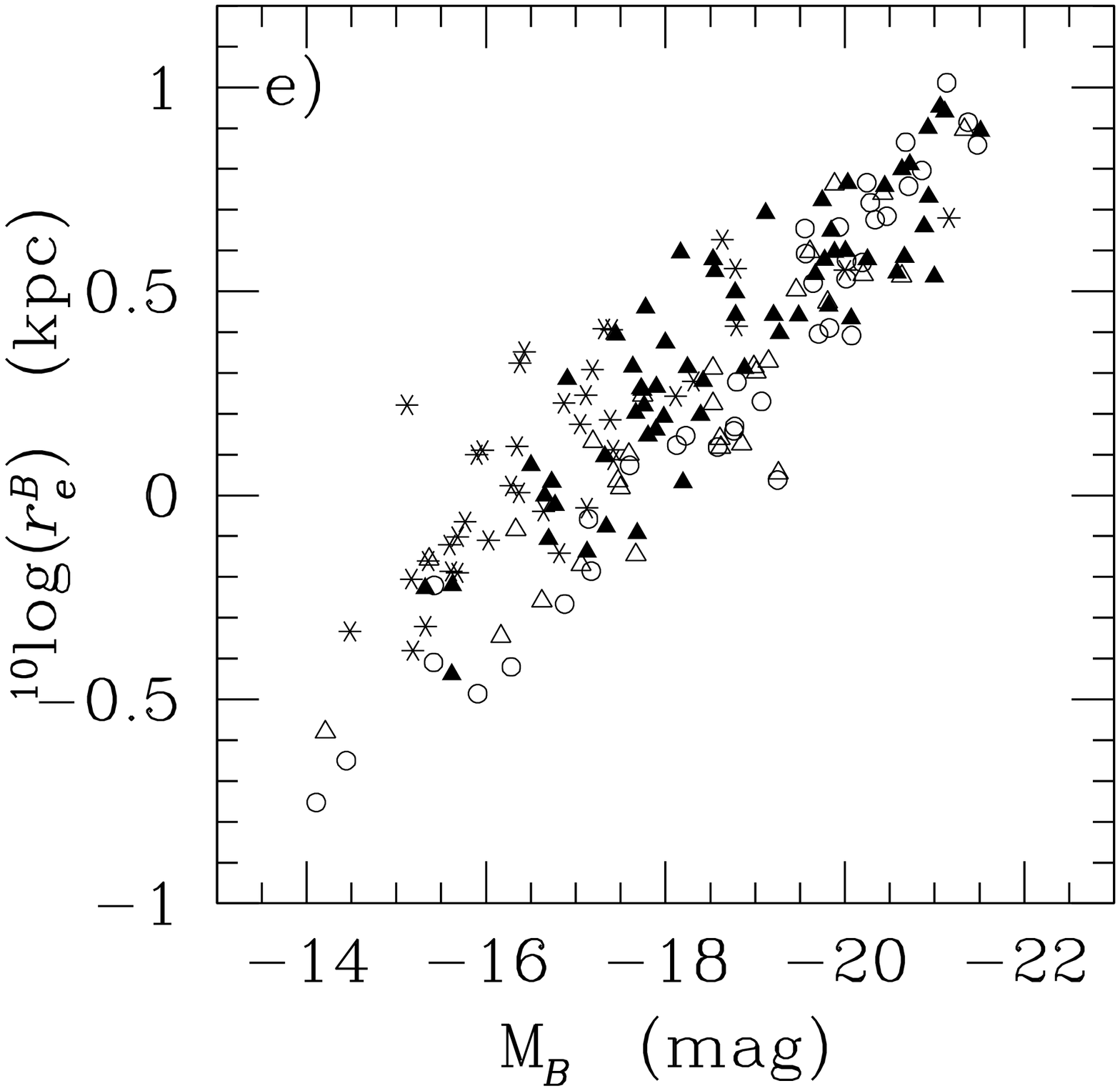,width=0.33\txw,clip=}
      \epsfig{file=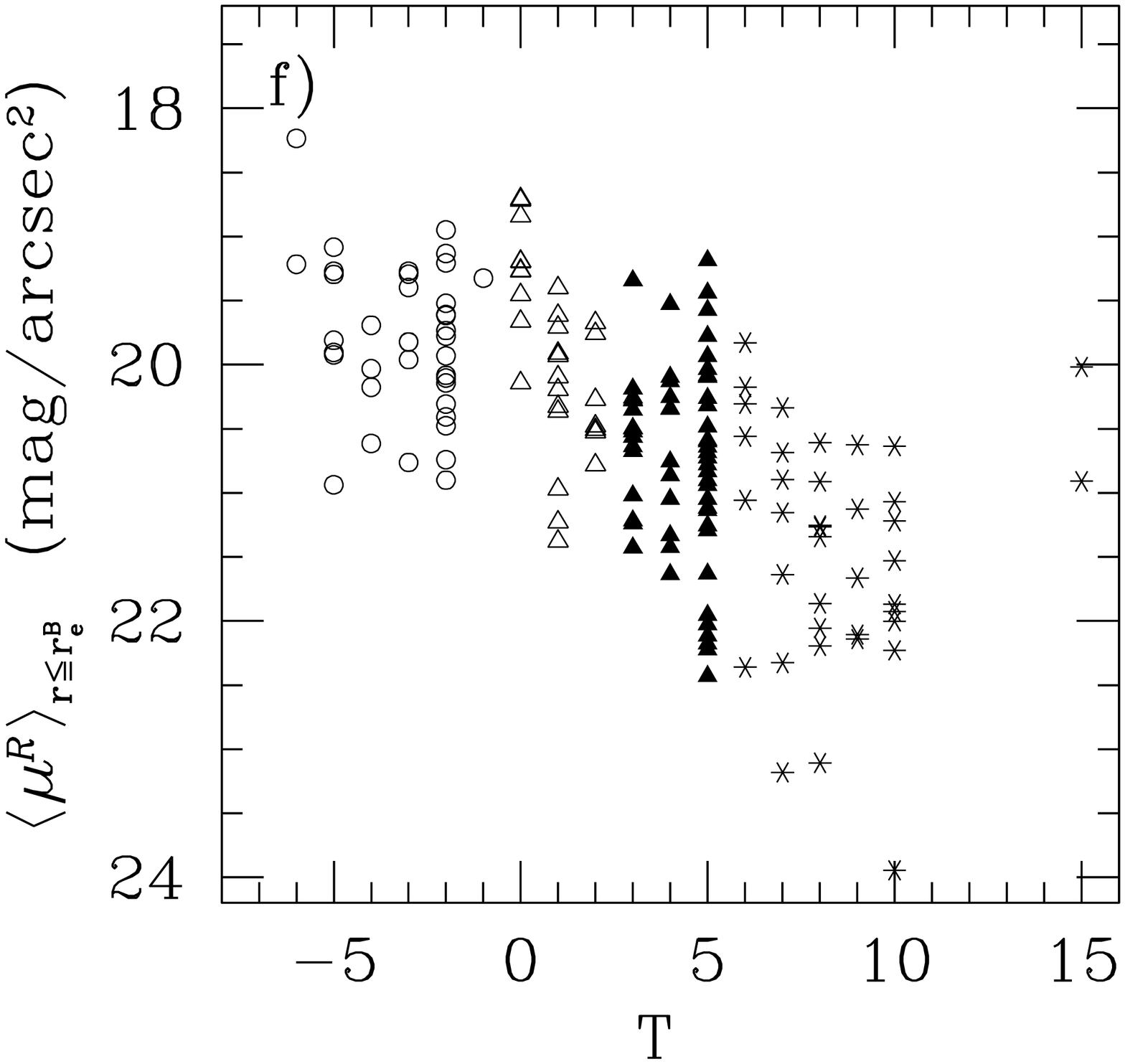,width=0.33\txw,clip=}
   }
}\par\noindent\leavevmode
\makebox[\txw]{
\centerline{
\parbox[t]{\txw}{\footnotesize {\sc Fig.~5 ---} Photometric properties
of the sample.  Points are coded according to type as follows.  For the
normal galaxies, ellipticals and S0's ($\mbox{T}<0$) are indicated by
open circles, early type spirals ($0\leq\mbox{T}<3$) by open triangles,
intermediate type spirals ($3\leq\mbox{T}<6$) by solid triangles and
late type spirals and irregulars ($\mbox{T}\geq 6$) by asterixes.  In
the first four panels, galaxies with active nuclei (AGN) are indicated
by crosses (Sy~{\sc i}, Sy~{\sc ii} and BL~Lac), starburst galaxies by
open stars, and the few galaxies with uncertain photometric zeropoint by
small symbols.\\ a) Total absolute $B$ filter magnitudes as a function
of numeric morphological type.  Apart from the group of faint early type
galaxies, the well-known trend of absolute magnitude with type --- later
type galaxies get progressively fainter --- is evident.  b) Effective
\BR\ colors as a function of type.  Galaxies become progressively bluer
towards later types.  c) Effective \BR\ colors versus total absolute $B$
magnitude.  Brighter galaxies tend to be redder.  d) Color--color plot
of effective \UB\ versus \BR\ colors.  Apart from the active ones, the
galaxies lie on a tight relation.  Note that this is the first such plot
for a sample so diverse in properties.  The interstellar extinction
vector is indicated.  e) Effective radii in $B$ versus total absolute
$B$ magnitude.  Apart from the (trivial) relation between galaxy size
and luminosity, we find a clear separation in type at a given magnitude
for $\MB>-19$.  Lower surface brightness galaxies lie above the mean
relation for a given absolute magnitude and given type.  f) Effective
$R$-filter surface brightnesses measured within the effective radii in
$B$ as a function of type.  Among the spiral galaxies a clear trend
towards fainter $\langle\mu^R\rangle_{r\leq r^B_e}$ going to later types
is visible.  } }
}\vspace*{0.5cm}

\noindent a reddening with radius are blue, but that many redder systems
shows this behaviour as well.  Panel {\em d} identifies these systems as
being physically smaller than average, while panel {\em a} shows that
there is no preference towards morphological type.  These results and
inspection of the images in the atlas (\eg IC~692 (E), A02257$-$\-0134
(Sdm), and A09125$+$\-5305 (Im)) suggest that star formation is the
driving force in this reddening-with-radius trend; in large, luminous
galaxies the star formation is stronger in the outer parts, while in
smaller, fainter systems star formation may occur anywhere.  In the
latter case, a single \HII-region can dominate the colors and its
location may determine the radial color trend.

%%%%%%%%%%%%%%%%%%%%%%%%%%%%%%%%%%%%%%%%%%%%%%%%%%%%%%%%%%%%%%%%%%%%%%%%
% SECTION  SECTION  SECTION  SECTION  SECTION  SECTION  SECTION  SECTION
%%%%%%%%%%%%%%%%%%%%%%%%%%%%%%%%%%%%%%%%%%%%%%%%%%%%%%%%%%%%%%%%%%%%%%%%

\null\vspace*{0.90\txw}
\section{Summary}
\label{P-discussion}

We have completed a photometric and spectrophotometric survey of \tsim
200 nearby galaxies of all morphological types and with a range of over
8 magnitudes in absolute $B$ magnitude, with the goal of studying the
variation in current star formation rates, star formation histories and
metallicities as a function of galaxy type and luminosity.  With this
survey we extend the work of Kennicutt (1992) to lower luminosity
systems.  The $U,B,R$ surface photometry will aid our interpretation of
the spectrophotometry. 

Here we have described the selection of our sample, the photometric
observations and our data reduction techniques.  We have presented our
photometric data, \vfill 

%%%%%%%%%%%%%%%%%%%%%%%%%%%%%%%%%%%%%%%%%%%%%%%%%%%%%%%%%%%%%%%%%%%%%%%%%
%      FIGURE 6: COLOR DIFFERENCES BETWEEN INNER AND OUTER PARTS	%
%%%%%%%%%%%%%%%%%%%%%%%%%%%%%%%%%%%%%%%%%%%%%%%%%%%%%%%%%%%%%%%%%%%%%%%%%
\noindent\leavevmode
\makebox[\txw]{
   \centerline{
      \epsfig{file=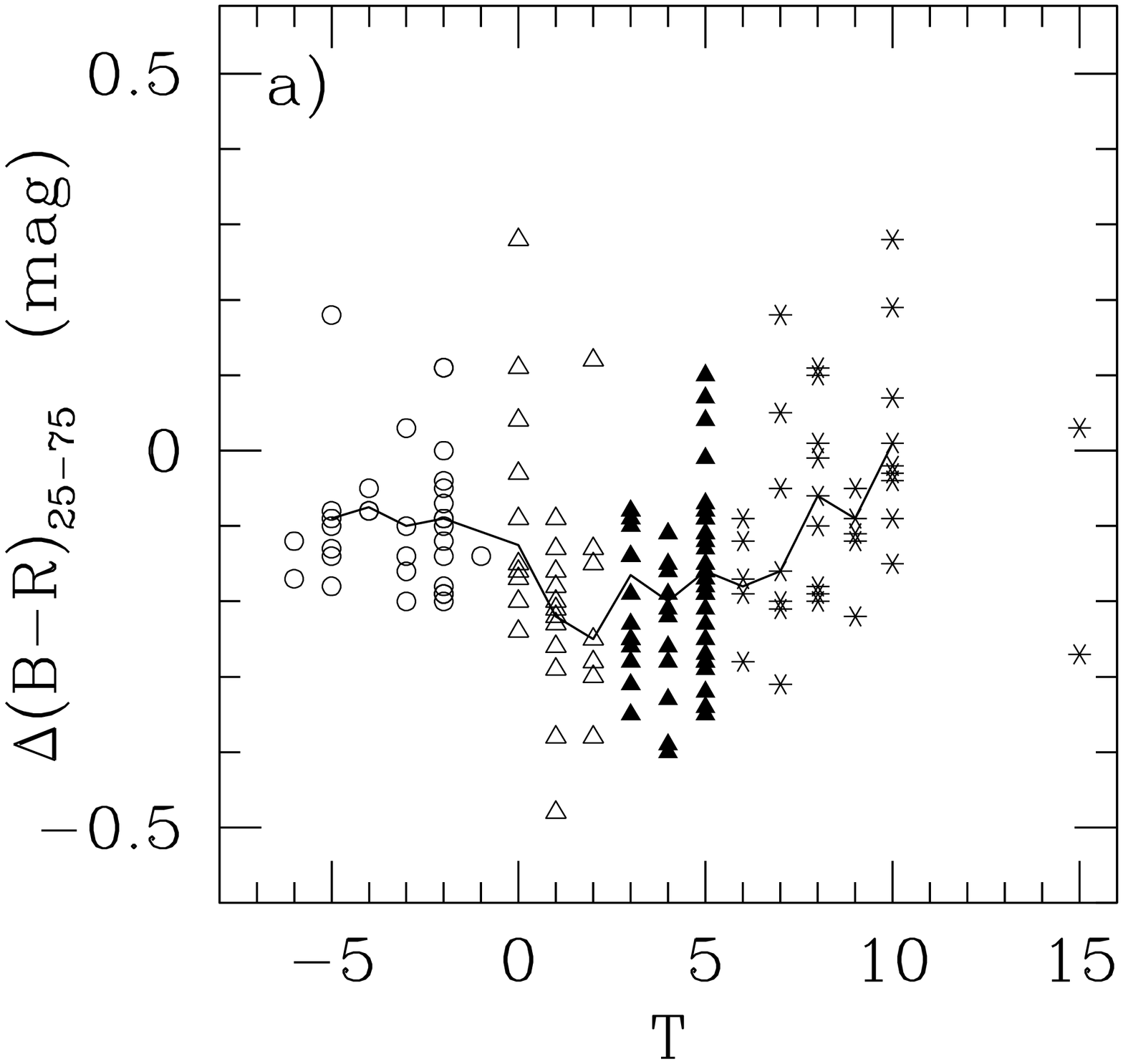,width=0.33\txw,clip=}
      \epsfig{file=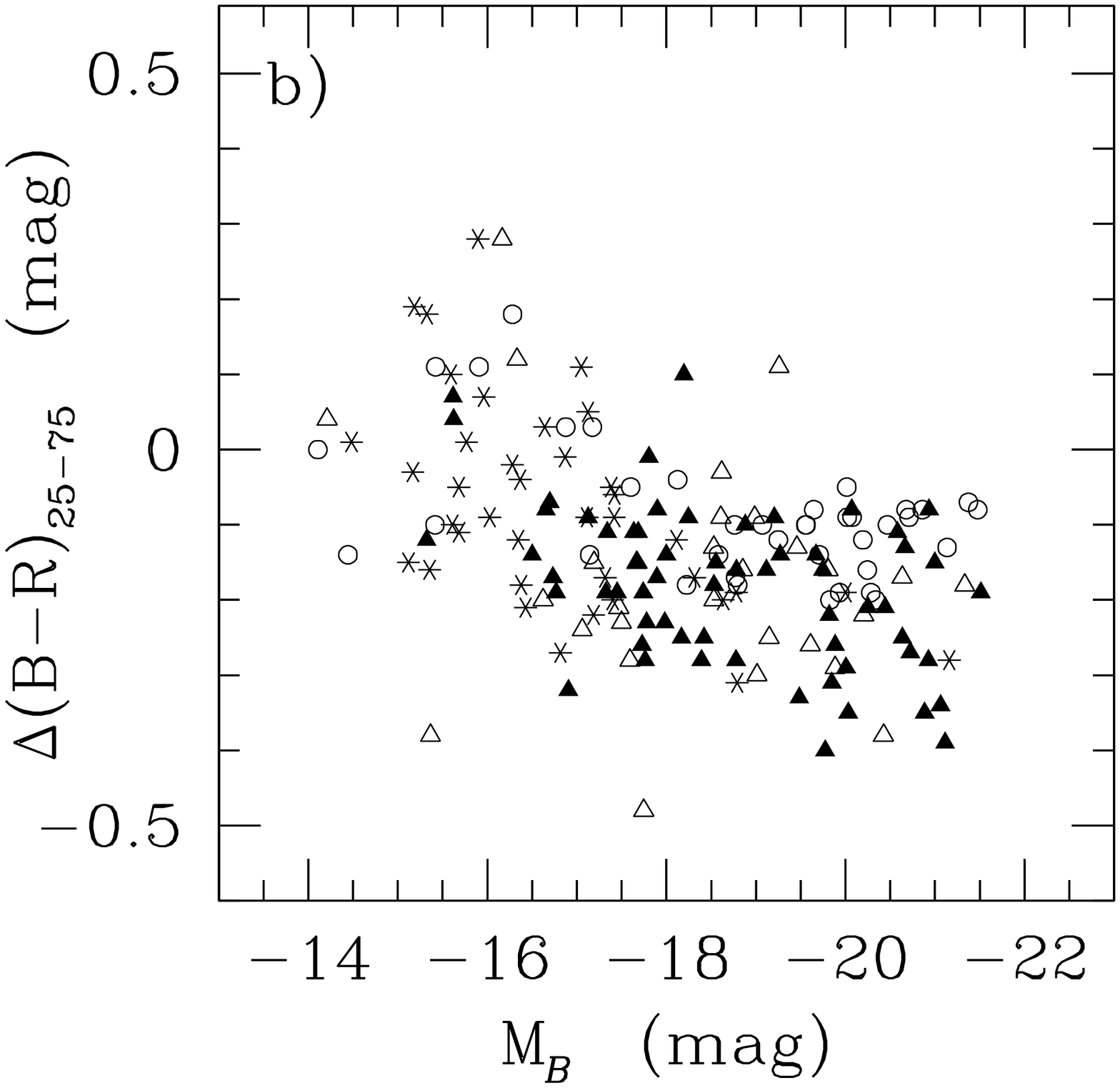,width=0.33\txw,clip=}
   }
}\par\noindent\leavevmode
\makebox[\txw]{
   \centerline{
      \epsfig{file=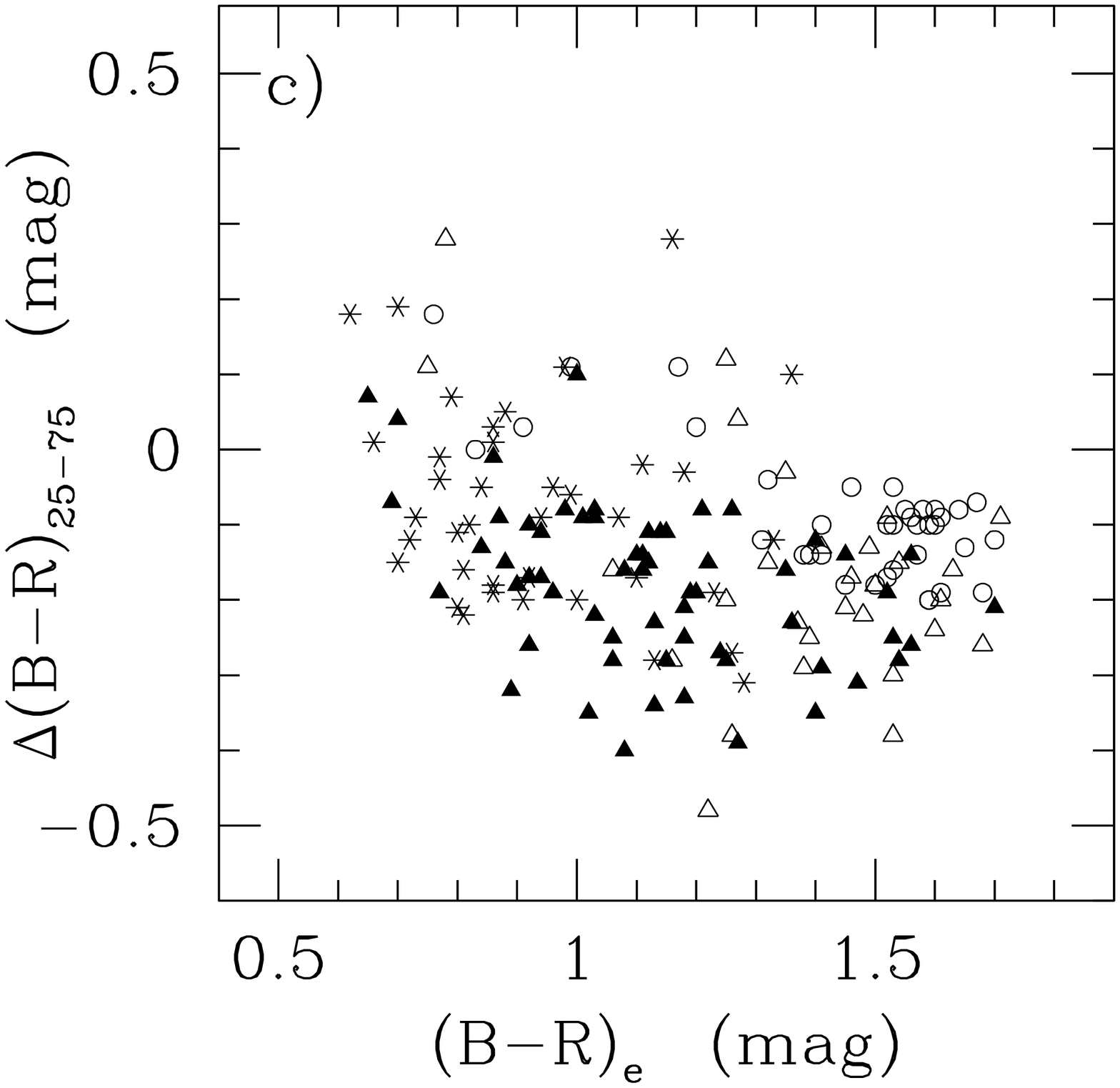,width=0.33\txw,clip=}
      \epsfig{file=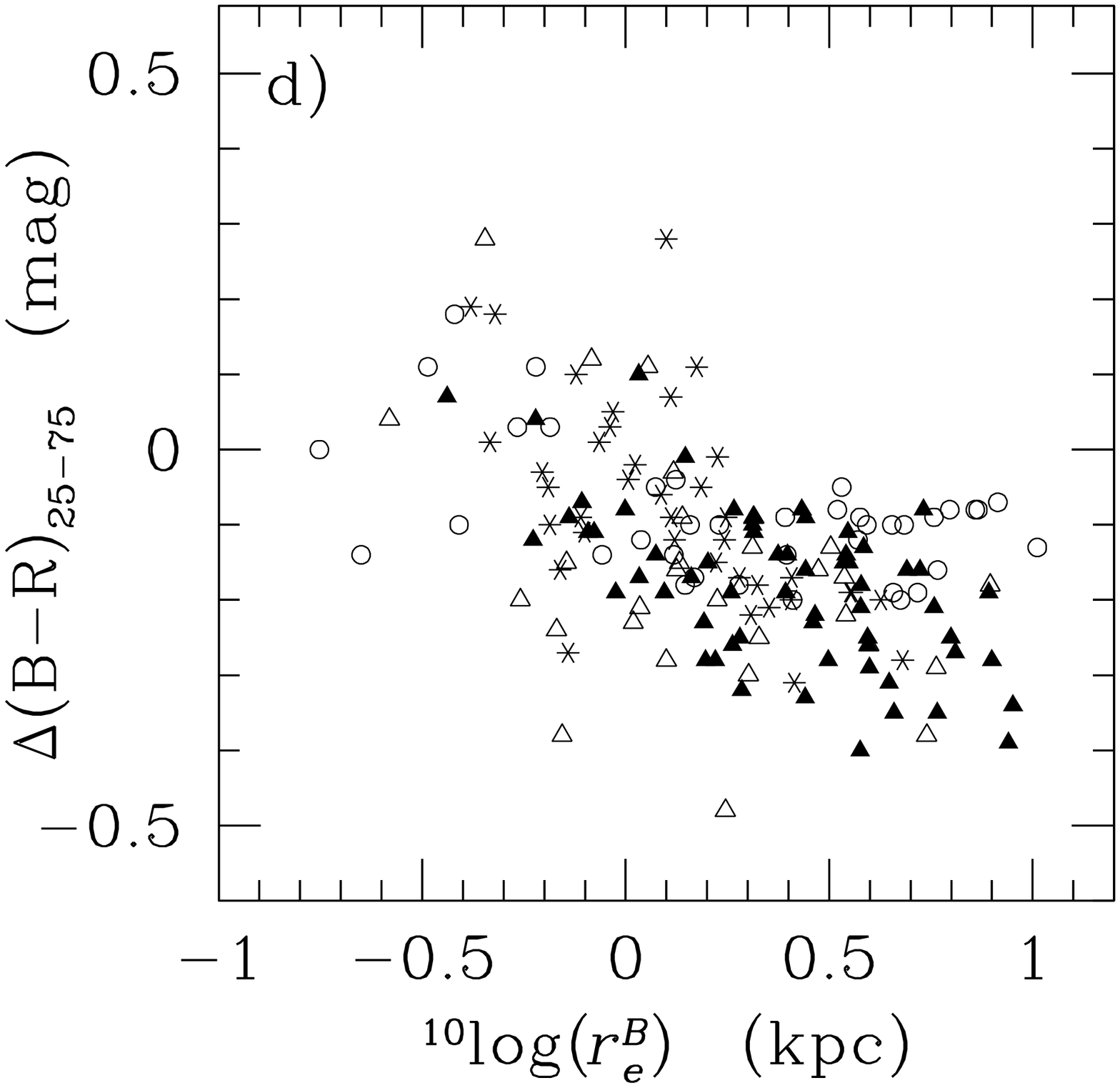,width=0.33\txw,clip=}
   }
}\par\noindent\leavevmode
\makebox[\txw]{
\centerline{
\parbox[t]{\txw}{\footnotesize {\sc Fig.~6 ---} Color differences
between inner and outer parts.  a) The color difference $\Delta
(B-R)_{25-75}$ (see text) as a function of type.  The median color
difference values have been connected by solid lines.  The disappearance
of a significant bulge going from early type spirals towards later types
causes the strong trend for T$>0$.  b) $\Delta (B-R)_{25-75}$ as a
function of total absolute $B$ magnitude.  Note that the galaxies
fainter than $\MB\sim -16$ are equally likely to become redder with
radius as they are to become bluer.  c) $\Delta (B-R)_{25-75}$ as a
function of effective \BR.  Only a small majority of the galaxies
showing a reddening trend with increasing radius is blue.  d) $\Delta
(B-R)_{25-75}$ as a function of effective radius in $B$.  Galaxies
showing a reddening trend with radius are intrinsically small systems. 
} } }\vspace*{0.5cm}

\noindent including an atlas of galaxy images with radial profiles of
surface brightness and \UB\ and \BR\ colors, and tables of photometric
measurements.  The spectrophotometry will be presented in a companion
paper.  Our surface photometry is consistent with previous published
work in almost all cases where a comparison is possible, and the
exceptions are described. 

We find a strong trend of \BRe\ color with morphological type, with
later type galaxies becoming progressively bluer.  The observed scatter
on this trend is 0.19 mag, which is smaller than the 0.24 mag scatter on
the color--magnitude trend where intrinsically fainter systems tend to
be bluer than brighter systems.  Estimating a galaxy's broad type class
(E,S,Irr) from its color can, therefore, be as accurate as estimates
based on galaxy asymmetry and central concentration (Abraham \etal
1996).  We find that color--magnitude relations are useful for early
type systems and to a lesser degree for very late type sys- \vfill

\null\vspace*{0.80\txw}\noindent tems, but not useful for intermediate
type spirals.  A color--color plot of \UBe\ versus \BRe\ shows a well
defined relation with small scatter, considering the wide range in
galaxy types, luminosities and colors in the sample. 

Galaxies brighter than $\MB=-17$ are almost always reddest in their
inner parts, but half the galaxies fainter than $\MB=-17$ are {\em
bluest} in their inner parts.  The total scatter in the radial color
difference is large (\tsim 0.5 mag).  Early and late type galaxies
behave similarly and we conclude that low luminosity dwarf galaxies
forming stars in their inner parts account for the reddening with
radius.  Tully \etal (1996) find that faint blue galaxies in the Ursa
Major cluster behave similarly.

\samepage {Both the average difference in color between the inner and
outer parts of a galaxy and the range in color difference from galaxy to
galaxy are smaller in early type galaxies than in spirals.}

%%%%%%%%%%%%%%%%%%%%%%%%%%%%%%%%%%%%%%%%%%%%%%%%%%%%%%%%%%%%%%%%%%%%%%%%
% ACKNOWLEDGEMENTS  ACKNOWLEDGEMENTS  ACKNOWLEDGEMENTS  ACKNOWLEDGEMENTS
%%%%%%%%%%%%%%%%%%%%%%%%%%%%%%%%%%%%%%%%%%%%%%%%%%%%%%%%%%%%%%%%%%%%%%%%
\onecolumn

%\acknowledgements
%\centerline{\bf Acknowledgements}
\section*{Acknowledgements}

\null\vspace*{0.1cm}
This work was supported by grants from the University of Groningen, the
Leiden Kerkhoven-Bosscha Fund, the Netherlands Organisation for
Scientific Research (NWO), and by the Smithsonian Institution.  RAJ
thanks the Harvard-Smithsonian Center for Astrophysics and the
F.L.~Whipple Observatory for hospitality during numerous visits, when
all of the observations and part of this work was done.  We thank
C.~Brice\~no, N.A.~Grogin, K.K.~McLeod, and T.~Groner for giving us some
of their observing time. We acknowledge comments made by the referee,
Dr.~G.D.~Bothun, which helped to clarify some aspects of this paper.

%%%%%%%%%%%%%%%%%%%%%%%%%%%%%%%%%%%%%%%%%%%%%%%%%%%%%%%%%%%%%%%%%%%%%%%%
% REFERENCES  REFERENCES  REFERENCES  REFERENCES  REFERENCES  REFERENCES
%%%%%%%%%%%%%%%%%%%%%%%%%%%%%%%%%%%%%%%%%%%%%%%%%%%%%%%%%%%%%%%%%%%%%%%%

%%%%%%%%%%%%%%%%%%%%%%%%%%%%%%%%%%%%%%%%%%%%%%%%%%%%%%%%%%%%%%%%%%%%%%%%
% APPENDIX  APPENDIX  APPENDIX  APPENDIX  APPENDIX  APPENDIX  APPENDIX
%%%%%%%%%%%%%%%%%%%%%%%%%%%%%%%%%%%%%%%%%%%%%%%%%%%%%%%%%%%%%%%%%%%%%%%%

\appendix
\section{A.\ \ Discussion of individual objects}
\label{P-App.objects}

\small

\noindent{\bf 006 A00510$+$1225} Compact Sc galaxy with Seyfert~{\sc i}
nucleus.  This galaxy was classified as compact elliptical in CfA~I. 

\noindent{\bf 007 NGC~315} Aka.  Holmberg~29A. 

\noindent{\bf 009 (A01047$+$1625)} This galaxy is not part of our
statistical sample.  It was observed because its estimated absolute
magnitude was $\sim -13$, making it a nice addition to our faintest bin
in absolute magnitude.  Recent observations by Hopp (1999), however,
suggest a much larger distance, and therefore higher luminosity
($\MB\sim -14.5$), than adopted in this paper. 

\noindent{\bf 010 NGC~382} Close companion of NGC~383 in this group of
galaxies. 

\noindent{\bf 011 IC~1639} Aka.  Markarian~562. 

\noindent{\bf 012 A01123$-$0046} Compact Sc galaxy previously classified
as compact elliptical in CfA~I.  This galaxy is interacting with two
close companions. 

\noindent{\bf 018 A01374$+$1539B} This is the lowest surface brightness
galaxy in our sample with a central surface brightness, $\mu^B_d(0)\sim
24.5$ mag\persecsq, and is also known as DDO~13. 

\noindent{\bf 027 NGC~927} Aka.  Markarian~593. 

\noindent{\bf 031 A02464$+$1807} is a compact galaxy dominated by a
foreground star right on its center.  We succeeded neither in obtaining
surface photometry of the underlying galaxy, nor spectroscopy for this
galaxy.  In the atlas we only show the image.

\noindent{\bf 032 A02493$-$0122} is another LSB galaxy.  Also known as
DDO~30. 

\noindent{\bf 041 NGC~2799} VV~50, Arp~283; interacting pair with
NGC~2798. 

\noindent{\bf 042 NGC~2824} Aka.  Markarian~394.  The light of a very
bright star \tsim 3\arcmin\ away from the galaxy needed to be modelled
out prior to the ellipse fitting step. 

\noindent{\bf 044 NGC~3011} Aka.  Markarian~409. 

\noindent{\bf 047 A09557$+$4758} Through the irregular spiral disk of
this galaxy, a more distant edge-on Sb is seen, giving the appearance of
a break (PA=42\arcdeg) in one of the arms of A09557$+$4758. 

\noindent{\bf 050 NGC~3104} This LSB without a distinct nucleus is also
known as VV~119. 

\noindent{\bf 052 NGC~3165} The central parts of this galaxy are
distinctly asymmetric. Possible companion to the Northeast.

\noindent{\bf 058 NGC~3279} Aka.  IC~622. 

\noindent{\bf 059 A10321$+$4649} Aka.  Markarian~146. Disturbed
morphology with a dust lane.

\noindent{\bf 062 A10365$+$4812} Galaxy showing the signature of a
strongly warped disk.

\noindent{\bf 063 A10368$+$4811} The light distribution in this galaxy
is asymmetric.

\noindent{\bf 064 NGC~3326} Aka.  AKN~251. Very faint extended spiral
structure is visible in this galaxy. 

\noindent{\bf 065 A10389$+$3859} Three smaller disk systems appear to
accompany this galaxy.

\noindent{\bf 066 A10431$+$3514} An edge-on early-type disk galaxy is
seen through this Sa galaxy, 14\arcsec\ Northeast of the nucleus.

\noindent{\bf 072 NGC~3499} A dustlane is running over this S0/a galaxy.

\noindent{\bf 074 A11017$+$3828W} This is the nearest galaxy with a
BL~Lac type nucleus.  It is highly variable (observed changes of \tsim
0.02 mag per night during its high state in May 1996).  The photometry
presented in this paper is an average of observations from March 1995,
and March and May 1996.  Due to the brightness of this object (forcing
us to use very short exposures) and due to the presence of two bright
stars nearby (leading to inaccurate measurement of the sky background,
even after modelling out the stellar light), our photometry of the host
galaxy is not deep. 

\noindent{\bf 077 IC~673} Named IC~678 in CfA~I. Galaxy with a low
surface brightness disk and a normal surface brightness bulge.

\noindent{\bf 079 A11072$+$1302} Contains multiple bright knots;
peculiar galaxy, a likely merger.

\noindent{\bf 080 NGC~3605} Light of the larger companion galaxy
NGC~3607 needed to be modelled out prior to the ellipse fitting step. 

\noindent{\bf 088 A11336$+$5829} The positional angle listed in the UGC
has an incorrect sign and should be --9\arcdeg (or 171\arcdeg) rather
than 9\arcdeg. 

\noindent{\bf 089 NGC~3795A} Named A1136$+$5833 in CfA~I. 

\noindent{\bf 092 A11378$+$2840} Bright knots in the inner parts of an
otherwise smooth galaxy. 

\noindent{\bf 093 A11392$+$1615} Aka. AKN~311 and Holmberg~275. May form
a pair with a galaxy 2.3\arcmin\ to the West.

\noindent{\bf 094 NGC~3846} Highly asymmetric light distribution.
Outer parts look tidally disturbed.

\noindent{\bf 097 NGC~3913} Aka.  IC~740. The faint spiral structure
appears to wind both clock- and counterclockwise.

\noindent{\bf 105 A12001$+$6439} Aka. Markarian~195.

\noindent{\bf 116 NGC~4288} Aka.  DDO~119. 

\noindent{\bf 119 A12195$+$7535} Seyfert~{\sc i} galaxy Markarian~205,
classified as compact elliptical, seen through the spiral disk of
foreground galaxy NGC 4319.

\noindent{\bf 120 A12263$+$4331} Aka. D~129.

\noindent{\bf 130 NGC~4841B} Fainter of the two galaxies, to the
northeast of brighter galaxy NGC~4841A. These two galaxies have a large
overlap on the sky, resulting in shallow photometry.

\noindent{\bf 151 NGC~5491} An apparent smaller companion is located
25\arcsec\ North of the nucleus.

\noindent{\bf 152 NGC~5532} A smaller companion galaxy, 0.5\arcmin\ 
South of the center of this elliptical, is overlapping the outer parts.

\noindent{\bf 153 NGC~5541} Aka.  AKN~444. 

\noindent{\bf 154 NGC~5596} Aka.  Markarian~470. 

\noindent{\bf 158 NGC~5762} This galaxy harbors a very extended and low
surface brightness flocculent disk.

\noindent{\bf 159 A14489$+$3547} This galaxy shows a highly warped,
banana-shaped disk and forms a pair with A14492$+$3545.  Also known as
II~Zw~70 and VV~324. 

\noindent{\bf 160 A14492$+$3545} Forms a pair with A14489$+$3547 (see
remark above) and shows the morphological signature of a polar ring. 
Also known as II~Zw~71 and VV~324. 

\noindent{\bf 161 IC~1066} A large barred spiral galaxy is located
2.1\arcmin\ North of this galaxy.

\noindent{\bf 163 A15016$+$1037} Seyfert~{\sc i} nucleus. Also known as
Mar\-ka\-rian 841.

\noindent{\bf 166 NGC~5875A} Named A1507$+$5229 in CfA~I. 

\noindent{\bf 169 NGC~5940} Galaxy with a weak Seyfert~{\sc i} nucleus;
also known as Markarian~9030. 

\noindent{\bf 171 NGC~5993} Two stars are superposed on this galaxy on
either side of the nucleus.  This galaxy as well as its neighbour
2.4\arcmin\ to the SW (not in this sample) show hints of tidal features
and are probably in interaction. 

\noindent{\bf 173 IC~1144} Aka.  Markarian~491.

\noindent{\bf 175 A15523$+$1645} Aka. AKN~489. 

\noindent{\bf 181 NGC~7077} Aka.  AKN~549. 

\noindent{\bf 183 A22306$+$0750} Also known as II~Zw181 and AKN~558.

\noindent{\bf 193 NGC~7620} Aka.  Markarian~321. 

\noindent{\bf 194 A23264$+$1703} Harbors a double nucleus and shows
tidal features of an ongoing interaction with a smaller companion. 

\noindent{\bf 196 NGC~7752} Strongly interacting companion of NGC 7753
with one of the tidally disrupted arms of NGC 7753 causing a starburst
on one side of NGC 7752.  Also known as AKN 585, IV Zw165 and VV~5. 

\noindent{\bf 197 A23514$+$2813} Named NGC~7777 in CfA~I.

\noindent{\bf --- (NGC~4319)} Foreground spiral galaxy through which
A12195$+$7535 is seen.

%%%%%%%%%%%%%%%%%%%%%%%%%%%%%%%%%%%%%%%%%%%%%%%%%%%%%%%%%%%%%%%%%%%%%%%%
% APPENDIX  APPENDIX  APPENDIX  APPENDIX  APPENDIX  APPENDIX  APPENDIX
%%%%%%%%%%%%%%%%%%%%%%%%%%%%%%%%%%%%%%%%%%%%%%%%%%%%%%%%%%%%%%%%%%%%%%%%

\section{B.\ \ Calibration of the intensity profiles}
\label{P-App.calib}

\small

We photometrically calibrated the radial intensity profiles by solving
simultaneously for the calibration equations for $B$ and $R$, or $U$ and
$R$, using \BR\ and \UR\ color terms, respectively. Consider the
pair of calibration equations for the $B$ and $R$ intensity profiles:
\begin{eqnarray}
   \left\{
      \begin{array}{rcl}
	\mu_B(r) &=& \mu_{B,i}(r) + pz_1 - pe_1\cdot X_1
			- pc_1\cdot (\mu_B-\mu_R)(r) \\
	\mu_R(r) &=& \mu_{R,i}(r) + pz_2 - pe_2\cdot X_2
			- pc_2\cdot (\mu_B-\mu_R)(r) \\
      \end{array}
   \right. \quad ,
\label{P-Eqn.photeqn}
\end{eqnarray}
where $\mu_{\lambda ,i}$ denotes the surface brightnesses in $B$ and $R$
in instrumental mag\persecsq, $pz_n$ denote the photometric zeropoints,
$pe_n$ the specific atmospheric extinction coefficients and $X_n$ the
airmass at the middle of the exposures, and $pc_n$ denote the color term
coefficients corresponding to our choice of reference filters. 

If we denote by $P_n(r)$ the terms that are independent of color,
$\mu_{\lambda ,i}(r)+pz_n-pe_n\cdot X_n$, we can rewrite
equations~(\ref{P-Eqn.photeqn}) as
\begin{eqnarray}
   \left\{
      \begin{array}{rcl}
        (1+pc_1)\cdot \mu_B(r) &=& P_1(r) + pc_1\cdot \mu_R(r) \\
        (1-pc_2)\cdot \mu_R(r) &=& P_2(r) - pc_2\cdot \mu_B(r) \\
      \end{array}
   \right.\quad ,
\end{eqnarray}
from which we find after substitution of each of these equations
into each other:
\begin{equation}
  \mu_B(r) = \frac{(1-pc_2)\cdot P_1 + pc_1\cdot P_2}{1 + pc_1 - pc_2}
  \qquad \hbox{and} \qquad
  \mu_R(r) = \frac{(1+pc_1)\cdot P_2 - pc_2\cdot P_1}{1 + pc_1 - pc_2}\ \ .
\label{P-Eqn.photcalib}
\end{equation}
Equations~(\ref{P-Eqn.photcalib}) represent the solutions for the surface
brightness profiles, corrected for the color term, in each of the two
filters.  Similar equations hold for $U$ and $R$ and corresponding color
\UR.  This procedure avoids the introduction of additional
correlated errors in the radial surface brightness profiles due to the
two filters used in the color term correction, that one expects when
propagating errors in the more commonly applied iterative methods. 

For completeness we give also the equations used to find the formal
random errors on the surface brightness profiles thus derived:
\begin{eqnarray}
   \left\{
      \begin{array}{rcl}
	nm &=& 1 + pc_1 - pc_2 \\
	\sigma_{\mu_B}(r) &=& \frac{1}{nm}\cdot \sqrt{
	   (1-pc_2)^2\sigma_{P_1}^2 + pc_1^2\sigma_{P_2}^2 + 
	   (\mu^2_B(r)+P_1^2)\cdot\sigma_{pc_2}^2 +
	   (\mu^2_B(r)+P_2^2)\cdot\sigma_{pc_1}^2 } \\
	\sigma_{\mu_R}(r) &=& \frac{1}{nm}\cdot \sqrt{
	   (1+pc_1)^2\sigma_{P_2}^2 + pc_2^2\sigma_{P_1}^2 + 
	   (\mu^2_R(r)+P_1^2)\cdot\sigma_{pc_2}^2 +
	   (\mu^2_R(r)+P_2^2)\cdot\sigma_{pc_1}^2 } \\
      \end{array}
   \right. .
\end{eqnarray}
The third and fourth terms under the root-sign are the dominant ones. 
We assume here that the individual errors, $\sigma $, are Gaussian in
nature, and un-correlated.  This approximately holds at the level of
accuracy we are aiming for in this work. 

For the non-photometric profiles that were scaled to photometric
intensity levels, we use the calibration constants, airmass and pixel
scale of the photometric reference profile.  The error on the fitted
scale factor is included in quadrature in the error propagation.  

All unique $B$,$R$-filter pairs of intensity profiles were calibrated as
described above, after which multiple resulting surface brightness
profiles were combined into a single profile per filter.  In the $U$
calibration we paired all $U$ filter intensity profiles with a
corresponding $R$ profile, but only solved for the $U$ filter part of
equations~(\ref{P-Eqn.photcalib}), not the $R$ filter one, as the $B$
filter profiles had better S/N in general and were therefore better
suited for this purpose.

\normalsize

%%%%%%%%%%%%%%%%%%%%%%%%%%%%%%%%%%%%%%%%%%%%%%%%%%%%%%%%%%%%%%%%%%%%%%%%%
%                                                                       %
%                         PHOTOMETRY DATA PAPER                         %
%	   TABLE 3: PHOTOMETRIC MEASUREMENTS AND PARAMETERS		%
%                                                                       %
%%%%%%%%%%%%%%%%%%%%%%%%%%%%%%%%%%%%%%%%%%%%%%%%%%%%%%%%%%%%%%%%%%%%%%%%%
%\landscape
\setlength{\tabcolsep}{1pt}
%\scriptsize

%\endlandscape

%%%%%%%%%%%%%%%%%%%%%%%%%%%%%%%%%%%%%%%%%%%%%%%%%%%%%%%%%%%%%%%%%%%%%%%%

\end{document}